\documentclass[preprint,12pt,3p]{elsarticle}

\usepackage{matlab-prettifier}
\usepackage{graphicx}
\usepackage{makecell}
\usepackage{xfrac}
\usepackage{parskip}
\usepackage{adjustbox}
\usepackage{setspace}
\usepackage{lscape}
\usepackage[backref=page]{hyperref}
\hypersetup{ 
backref=true, % link backreferences (i.e. citations to their list)
bookmarks=true, % show bookmarks bar?
pdftoolbar=true, % show Acrobats toolbar?
pdfmenubar=true, % show Acrobats menu?
pdffitwindow=false, % window fit to page when opened
pdfstartview={FitH}, % fits the width of the page to the window
pdftitle={}, % title
pdfauthor={}, % author
pdfsubject={}, % subject of the document
pdfkeywords={}{}, % list of keywords
pdfnewwindow=true, % links in new window
colorlinks=true, % false: boxed links; true: colored links
linkcolor=BlueViolet, % colour of internal links
citecolor=maroon, % colour of links to bibliography
urlcolor=blue,
}
\usepackage{comment}
\usepackage{subfig}
\usepackage{caption}
\usepackage{longtable}
\usepackage{amssymb}
\usepackage{amsmath}
\usepackage{gensymb}
\usepackage{siunitx}

\usepackage{natbib}
\biboptions{comma,round}
\setcitestyle{authoryear}

\begin{document}
%\begin{spacing}{1.5}
\begin{frontmatter}

\title{Liabilities for the social cost of carbon}

\author[1,2,3]{Matthew K. Agarwala}
\author[4,5,6,7,8,9]{Richard S.J. Tol\corref{cor1}%\fnref{ack}
}
\address[1]{Bennett Institute for Innovation \& Policy Acceleration, University of Sussex, Falmer, United Kingdom}
\address[2]{Tobin Center for Economic Policy, Yale University, New Haven, CT, USA}
\address[3]{Bennett School of Public Policy, University of Cambridge, Cambridge, United Kingdom}
\address[4]{Department of Economics, University of Sussex, Falmer, United Kingdom}
\address[5]{Institute for Environmental Studies, Vrije Universiteit, Amsterdam, The Netherlands}
\address[6]{Department of Spatial Economics, Vrije Universiteit, Amsterdam, The Netherlands}
\address[7]{Tinbergen Institute, Amsterdam, The Netherlands}
\address[8]{CESifo, Munich, Germany}
\address[9]{Payne Institute for Public Policy, Colorado School of Mines, Golden, CO, USA}

\cortext[cor1]{Jubilee Building, BN1 9SL, UK}
%\fntext[ack]{}

\ead{r.tol@sussex.ac.uk}
\ead[url]{http://www.ae-info.org/ae/Member/Tol\_Richard}

\begin{abstract}
We estimate the national social cost of carbon using a recent meta-analysis of the total impact of climate change and a standard integrated assessment model. The average social cost of carbon closely follows per capita income, the national social cost of carbon the size of the population. The national social cost of carbon measures self-harm. Net liability is defined as the harm done by a country's emissions on other countries minus the harm done to a country by other countries' emissions. Net liability is positive in middle-income, carbon-intensive countries. Poor and rich countries would be compensated because their current emissions are relatively low, poor countries additionally because they are vulnerable.
\end{abstract}

\end{frontmatter}

\section{Introduction}
The social cost of carbon is the damage done by emitting an extra tonne of carbon dioxide. It is often conceptualized as the Pigou tax a philosopher-queen-of-the-world would impose. It can also be seen as the carbon tax in a cooperative game. In a non-cooperative game, where national planners ignore the damage outside their jurisdictions, the \emph{national} social cost of carbon is the relevant concept. It is the harm done by a country to itself. National social costs of carbon are (much) lower than global social costs: Most damages from domestic emissions accrue abroad. The U.S. Environmental Protection Agency (EPA) adopted this approach during President Trump’s first term. Alongside higher discount rates, this reduced the official U.S. estimate from \$43 to \$3–5 per tC. In  Trump’s second term, Executive Memorandum M-25-27 \citep{Clark2025} instructed federal agencies to ignore the social cost of carbon unless explicitly required by statute. At the time of writing, the EPA is “revisiting” the official estimate, including "consideration of eliminating the 'social cost of carbon' calculation from any Federal permitting or regulatory decision" \citep{Trump2025}.  This paper presents new estimates of the national social cost of carbon, discusses its key sensitivities and patterns, and applies the new numbers to estimate the net climate liabilities of almost every country.

The paper updates \citet{Tol2019EE}, which combined a meta-analysis of the total impact of climate change with a simple integrated assessment model to estimate the national cost of carbon. We here use data from a later meta-analysis \citep{Tol2024EnPol}. Data and insights on the total impact of climate change did not change much between these two papers; the social cost of carbon is therefore not very different either. The main innovations in the current paper are the blame matrices and the quantification of net liabilities.

\citet{Ricke2018} is the first paper to estimate the social cost of carbon at the national level. It unfortunately relies on the misspecified model by \citet[][cf. \citet{Newell2021}]{Burke2015}, which overstates the total and hence marginal impacts of climate change in hot countries and understates said impacts in cold countries. \citet{Chiani2026, Hwang2026, Rising2026} and \citet{Yoo2026} report new sets of estimates of the social cost of carbon; see \citet{Yoo2026summ} for a comparison. We further contribute to that by discussing how these estimates vary over time.

Liability, however, is the main contribution. Climate change, and hence its impacts, depend on cumulative emissions. Accumulation over space is straightforward: It does not matter who or what emits carbon dioxide. Emissions from one country are perfect substitutes for emissions from another country. A country's responsibility is proportional to its share in global emissions.

Accumulation over time is not straightforward. Carbon cycle and climate are coupled, non-linear, dynamic systems. To a first approximation, global warming is linear in cumulative emissions \citep{Matthews2009} so that a country's liability for \emph{warming} is roughly proportional to its share in historic emissions. However, liability is for harm done. Impacts are non-linear in warming. Early warming may have been beneficial. Abundant energy and food\textemdash the main sources of carbon dioxide emissions\textemdash certainly are beneficial. Holding people accountable for something they did when they could not have known its consequences is fraught with legal and moral difficulties. Territorial control and jurisdiction have changed (throughout) history. See \citet{Allen2003, Tol2004, James2014, Vanhala2016, Boyd2017} and \citet{Setzer2019} for an extensive discussion.

We therefore take a simpler approach: Liability is for current emissions. Harm is quantified by the social cost of carbon, the net present value of the future net damages done by a slight increase in emissions. Net liability equals the harm done to others minus the harm done by others.

This short paper proceeds as follows. Section \ref{sc:methods} presents the methods, section \ref{sc:results} the results. Section \ref{sc:conclude} concludes.

\section{Methods}
\label{sc:methods}
The integrated assessment model is standard. Population follows an exogenous scenario. Economic output is based on a Solow-Swan growth model with a Cobb-Douglas production function. Energy is a derived demand, with energy intensity steadily improving at an exogenous rate. The carbon intensity too is a scenario. The atmospheric concentration of carbon dioxide is based on a Maier-Reimer-Hasselmann carbon cycle model. The temperature follows a Schneider-Thompson climate model. This is almost identical to Nordhaus' DICE model.

\begin{figure}[hp]
    \centering
    \caption{Economic impacts of climate change}
    \label{fig:impact}
    \includegraphics[width=1.0\linewidth]{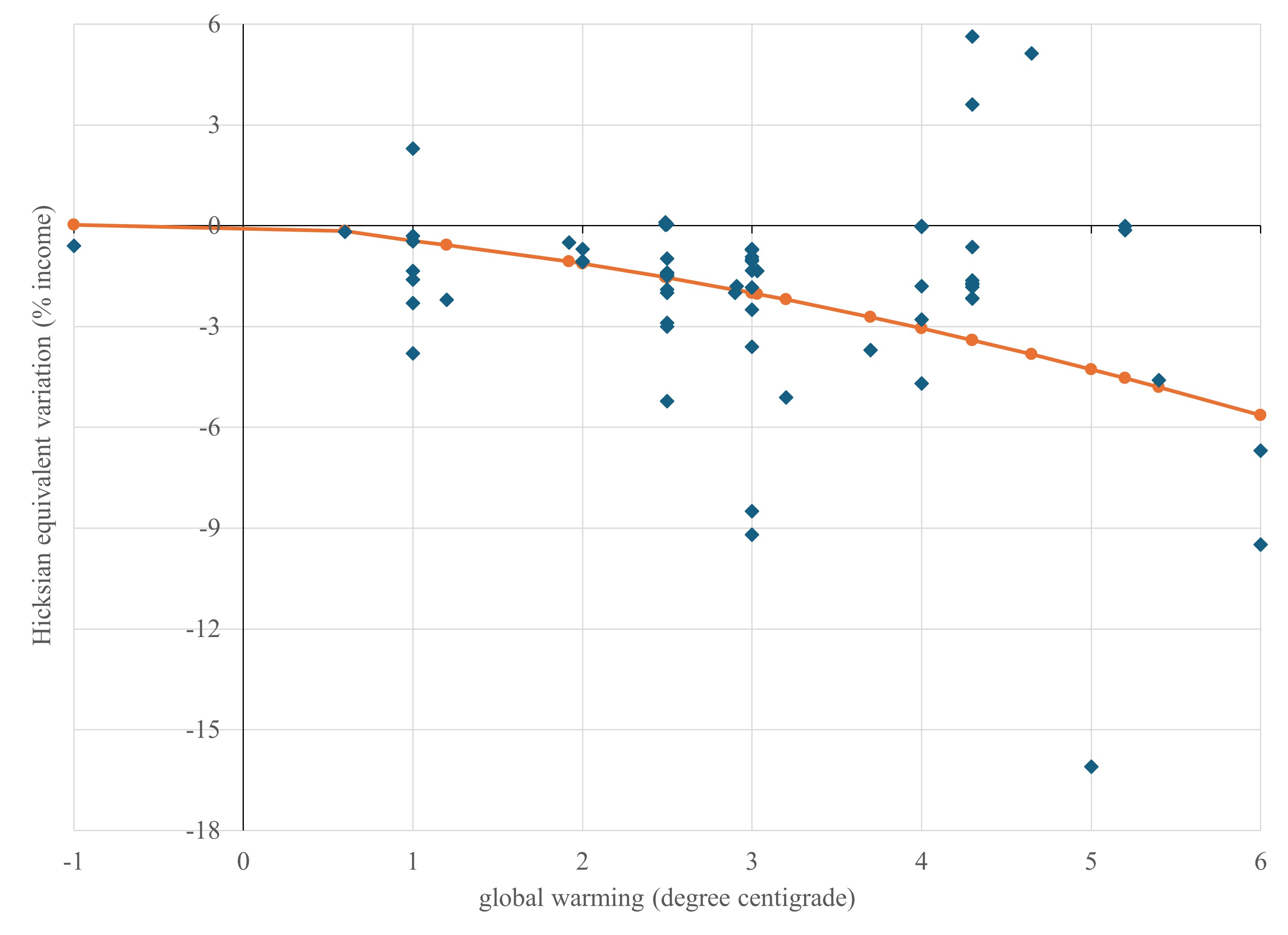}
    \caption*{\footnotesize The blue diamonds are the impacts of climate change reported in the meta-analysis of \citet{Tol2024EnPol}. The orange line in the Bayesian model average of the impact functions reported in Table \ref{tab:function}. The estimates are based on comparative statics, independent of time and economic development.}
\end{figure}

The impacts of climate change are different. \citet{Nordhaus1992} assumes a quadratic function, and has maintained this assumption until today \citep[][see \citet{NordhausMoffat2017} for an exception]{Barrage2024}. We instead consider nine alternative impact functions; see Table \ref{tab:function}. We estimate the parameters of these alternative functions using the data in \citet{Tol2024EnPol}. In the main analysis, we used the Bayesian Model Average of these functions. That is, we use the sum of squared residuals as the measure of model fit. Assuming Normality, we then compute the likelihood of each alternative. For comparison, we also include three high-pedigree parameterizations but, as shown in Table \ref{tab:function}, these are a poor fit. The default impact function is then the weighted average of the alternative functions, using their likelihoods as weights. See Figure \ref{fig:impact}.

There is a second difference. Where \citet{Nordhaus1992} and much of the literature on the social cost of carbon that followed in his footsteps assume that the impacts depend on climate change and climate change only, we follow \citet{Schelling1984, Schelling1992} and assume that richer countries are less vulnerable to climate change. Specifically, we follow \citet{Botzen2021} and adopt an income elasticity of -0.36.

The \emph{national} impact functions follow the imputation-after-calibration method proposed by \citet{Tol2019EE}. The parameters of the global impact functions are scaled to the national level using the income elasticity $\varepsilon$; and rescaled to add up to the global impact. The national impact estimates are assumed to be independent of one another, a convenient but flawed assumption \citep{Darwin2001, Feng2024}.
\begin{comment}
A Fehr-Schmidt utility function is often used to interpret behaviour in experiments:
\begin{equation}
    U(x_i) = x_i - \alpha \max_r (x_r - x_i, 0) - \beta \max_p (x_i - x_p, 0)
\end{equation}
where $\alpha$ measures \emph{disadvantageous aversion}, that is, a dislike for others being better off than you, and $\beta$ measures \emph{advantageous aversion}, that is, a dislike for others being worse off than you. The meta-analysis of \citet{Nunnari2022} finds $\alpha = 0.39$ and $\beta = 0.23$. A transfer of \$1 from $x_r$ to $x_p$ would improve utility by $\alpha + \beta$. This is the \emph{consumption} rate of inequity aversion.
\end{comment}

\section{Results}
\label{sc:results}
Figure \ref{fig:pop} plots the national social cost of carbon against the size of the population in 2010. The social cost of carbon is shown for the RFF scenario; there is no carbon tax in this scenario. The numbers can be found in Table \ref{tab:sccscen}. As more people means that more people suffer the impact of climate change, the relationship is positive: More populous countries have a larger social cost of carbon.

Figure \ref{fig:incelas} plots the national social cost of carbon against per capita income in 2010. As with population, a large economy implies that more damage is done. Comparing Figures \ref{fig:incelas} and \ref{fig:pop} shows that the relationship of the social cost of carbon with income is less noisy than the relationship with population.

Figure \ref{fig:incelas} shows three graphs. The first has the default assumption in \textsc{fund}: $\varepsilon<0$, poorer countries are \emph{more} vulnerable to climate change. This is the majority position in the literature on the impacts of climate change and natural disasters. The second has the default assumption in the literature on the social cost of carbon: $\varepsilon = 0$, rich and poor are equally vulnerable. The third follows the assumption in a small minority of studies \citep{Sterner2008, Drupp2021, Bremer2021}: $\varepsilon > 0$, poorer countries are \emph{less} vulnerable. Figure \ref{fig:incelas} shows that the slope of the relationship between per capita income and the national social cost of carbon depends on the assumed income elasticity $\varepsilon$.

Figure \ref{fig:carbon} plots the national social cost of carbon against carbon efficiency, that is, value added per tonne of carbon emitted. Total output is measured in the tens to trillions of dollars per year, total emissions in the tens of billions. Global average carbon efficiency is around \$6,000/tC. No relationship is apparent. However, carbon efficiency is always greater than the national social cost of carbon. This compares the \emph{average} value added to the \emph{marginal} external cost. Carbon efficiency varies strongly between economic sectors. Carbon \emph{productivity}, based on a production function including intermediates and thus the indirect economic effect of energy use, is a better measure than carbon \emph{efficiency}. These caveats notwithstanding, Figure \ref{fig:carbon} suggests that no country would risk economic ruin by imposing the \emph{national} social cost of carbon as a tax on its carbon dioxide emissions. Indeed, carbon pricing in the EU, at a level much higher than the national social costs of carbon discussed here, did not discernibly affect economic growth \citep[e.g.,][]{Metcalf2023}.

\begin{figure}[hp]
    \centering
    \caption{The external cost imposed \emph{by} each country plotted against the external cost imposed \emph{on} each country.}
    \label{fig:blame}
    \includegraphics[width=1.0\linewidth]{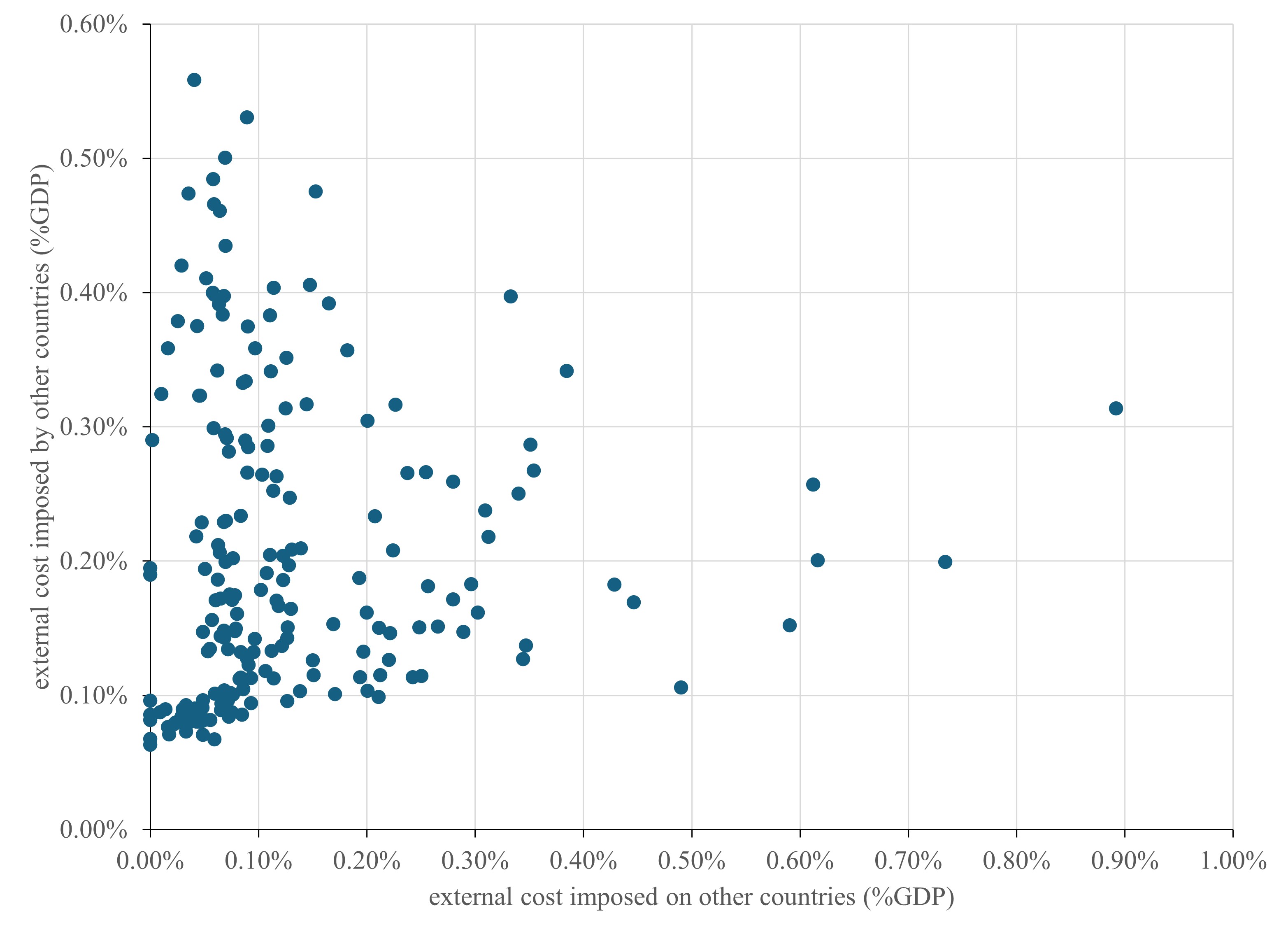}
    \caption*{\footnotesize Scenario: RFF; PRTP: 1.5\%; EIS: 1.5; impact: Bayesian model average; income elasticity of impact: -0.36; results are for 2010.}
\end{figure}

Figure \ref{fig:blame} shows the blame matrix. The horizontal axis has $H_c$ or \emph{harm done}, the externality posed by country $c$ \emph{on} other countries: The global social cost of carbon minus the national social cost of carbon times national emissions $M$:
\begin{equation}
    H_{c} = M_c \sum_{i \neq c} SCC_i 
\end{equation}
The vertical axis shows $D_c$ or \emph{damage suffered}, the externality posed \emph{by} other countries: The national social cost of carbon times global minus national emissions:
\begin{equation}
    D_{c} = SCC_c \sum_{i \neq c} M_i 
\end{equation}
Figure \ref{fig:blame} shows $H_c$ and $D_c$ normalized by GDP $Y_c$.

There is no obvious pattern, but some countries are at the receiving end while other countries cause substantial damage to others. Gross liability never exceeds 0.58\% of GDP.

Recall that these numbers are proportional to the \emph{global} social cost of carbon; the number used here, \$7/tC, is low compared to the literature \citep{Tol2023NCC}, primarily because we assume that relative impacts fall with economic growth \citep{Diaz2017}.

\begin{figure}[hp]
    \centering
    \caption{Net liability plotted against per capita income.}
    \label{fig:liability}
    \includegraphics[width=1.0\linewidth]{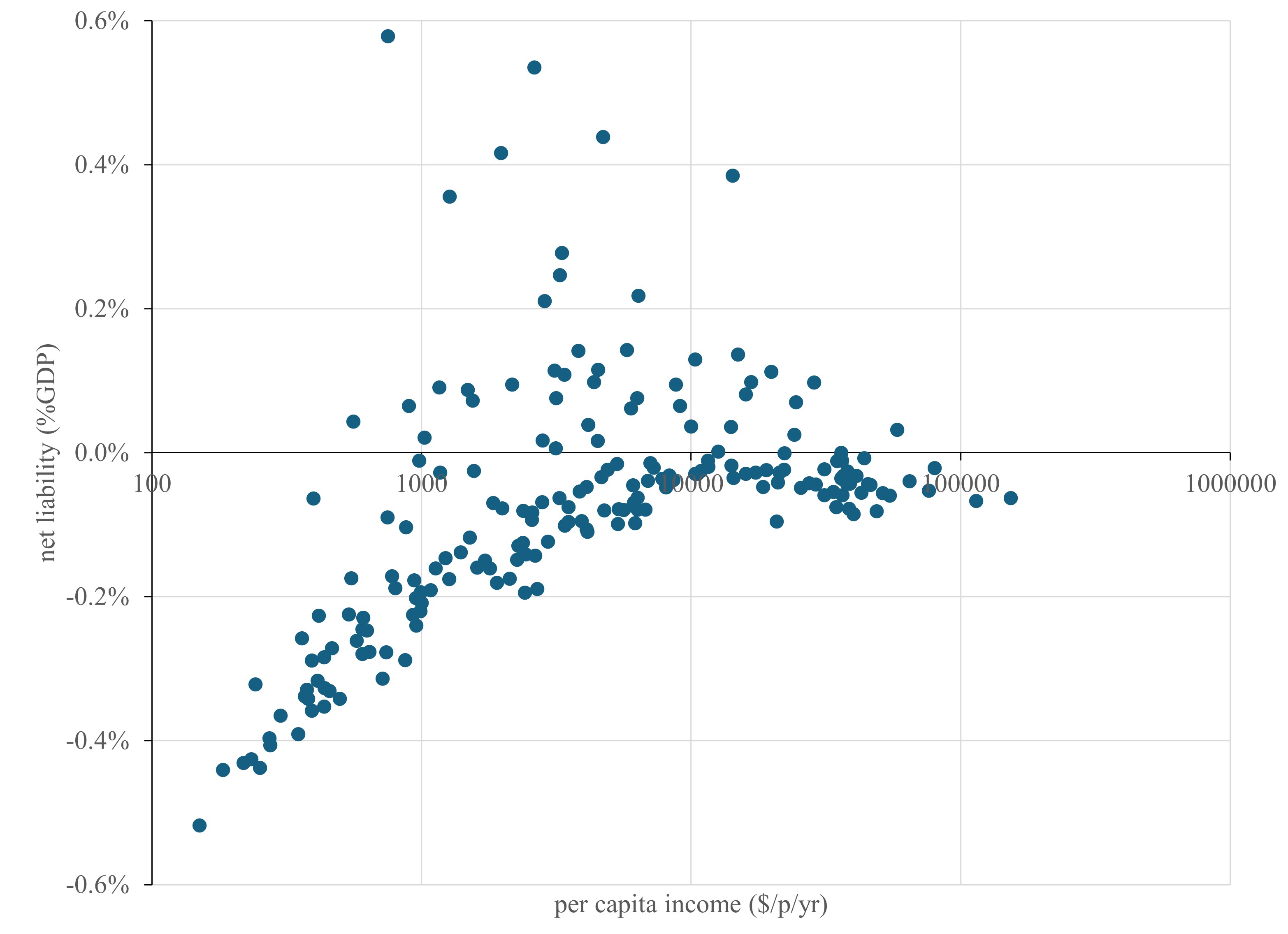}
    \caption*{\footnotesize Scenario: RFF; PRTP: 1.5\%; EIS: 1.5; impact: Bayesian model average; income elasticity of impact: -0.36; results are for 2010, dollars are 2005 U.S. dollars.}
\end{figure}

Figure \ref{fig:liability} plots \emph{net} liability against per capita income. Net liability $L_c$ is defined as the external cost posed on other countries (the horizontal axis in Figure \ref{fig:blame}; see also Figure \ref{fig:harm}) minus the external cost posed by other countries (the vertical axis in Figure \ref{fig:blame}; see also Figure \ref{fig:damage}):
\begin{multline}
\label{eq:netliability}
    L_c = H_c-D_c = \left ( M_c \sum_i SCC_i - SCC_c \sum_i M_i \right ) = \\
    \sum_i SCC_i \sum_i M_i \left ( \frac{M_c}{\sum_i M_i} - \frac{SCC_c}{\sum_i SCC_i} \right )
\end{multline}
Again, Figure \ref{fig:liability} shows net liability normalized by GDP $\sfrac{L_c}{Y_c}$.

Most countries have a negative net liability: They suffer greater damage than they cause harm to others. Countries with a positive net liability tend to be middle-income: The poorest countries emit little, the richest emit little relative to the size of their economy.

The second line of Equation (\ref{eq:netliability}) clarifies this. The sign of net liability is determined by the \emph{share} in global emissions minus the \emph{share} in the social cost of carbon. Poor countries contribute little to global emissions but a lot to the impacts of climate change. Rich countries are relatively less vulnerable to climate change but damages are large in absolute terms. These service-based economies are carbon-extensive so their share in global emissions is relatively small. See Figure \ref{fig:share}. As a result, carbon-intensive middle-income countries compensate the rest for the impacts of climate change.

\subsection{Sensitivity analysis}
The social cost of carbon depends on a great number of assumptions. Tables \ref{tab:sccscen} and \ref{tab:sccsens} show a limited sensitivity analysis. Below I highlight the impact on net liability.

\begin{figure}[hp]
    \centering
    \caption{Net liability: Sensitivity analysis.}
    \label{fig:liasens}
    \includegraphics[width=0.49\linewidth]{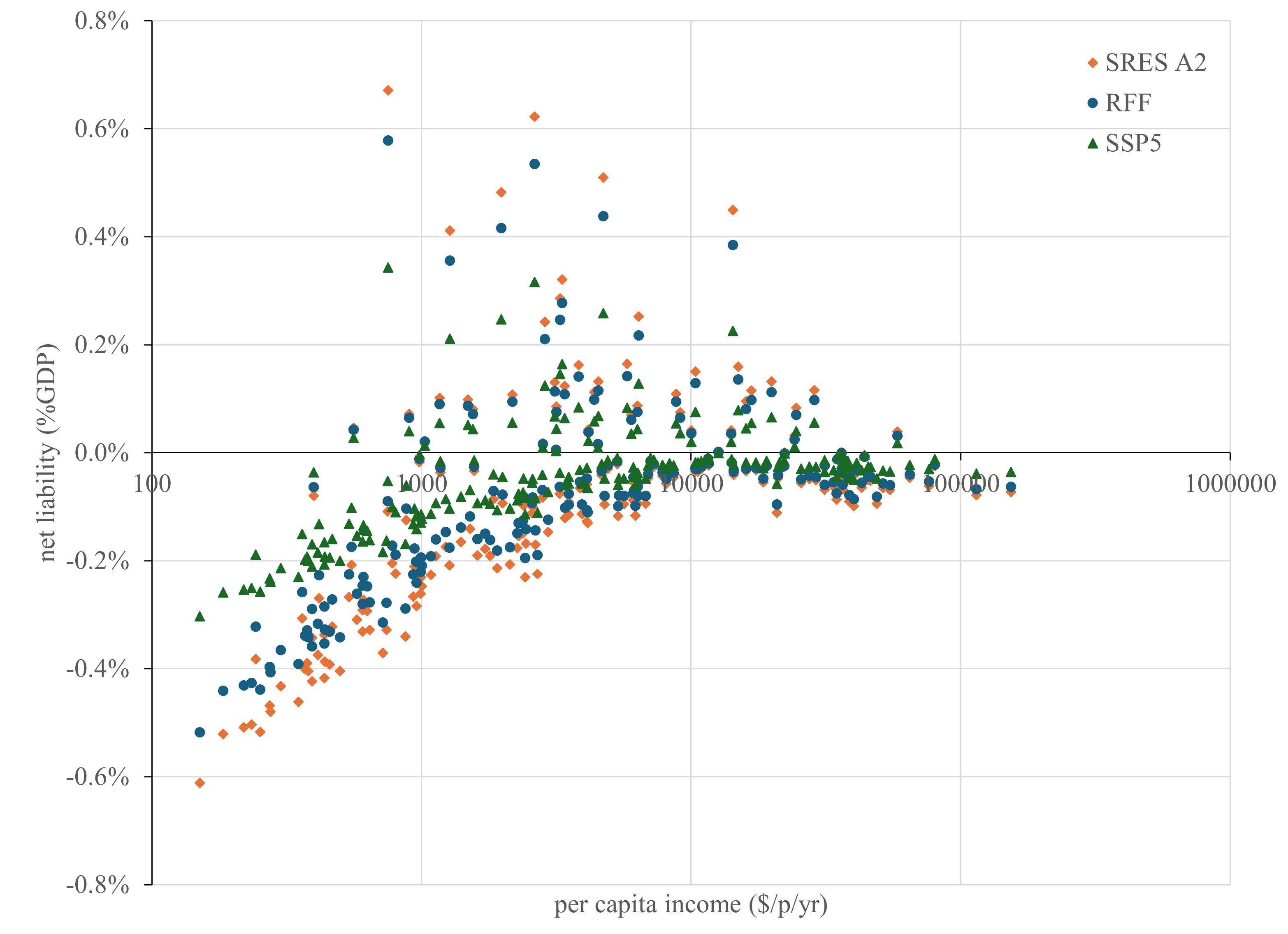}
    \includegraphics[width=0.49\linewidth]{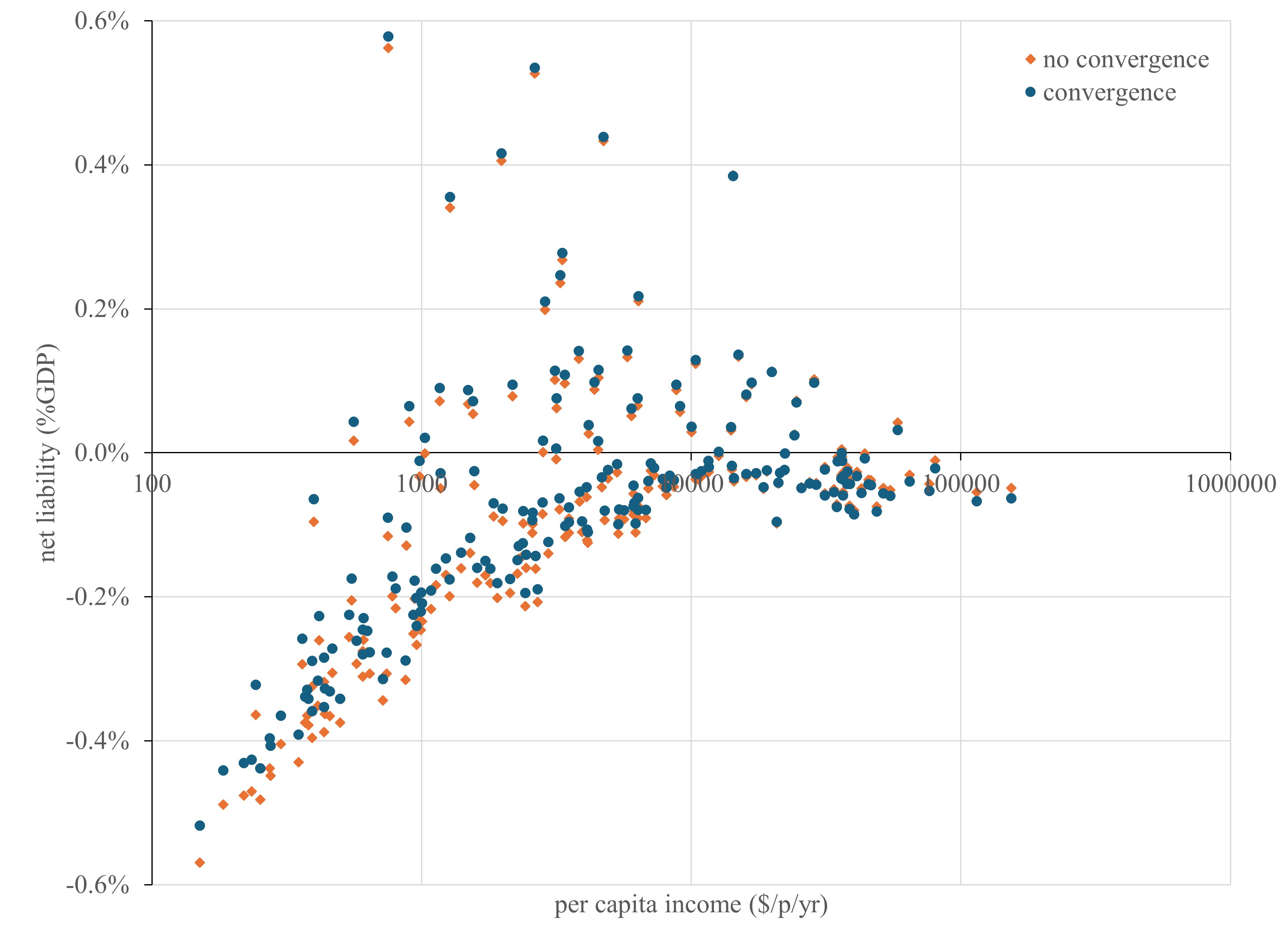}
    \includegraphics[width=0.49\linewidth]{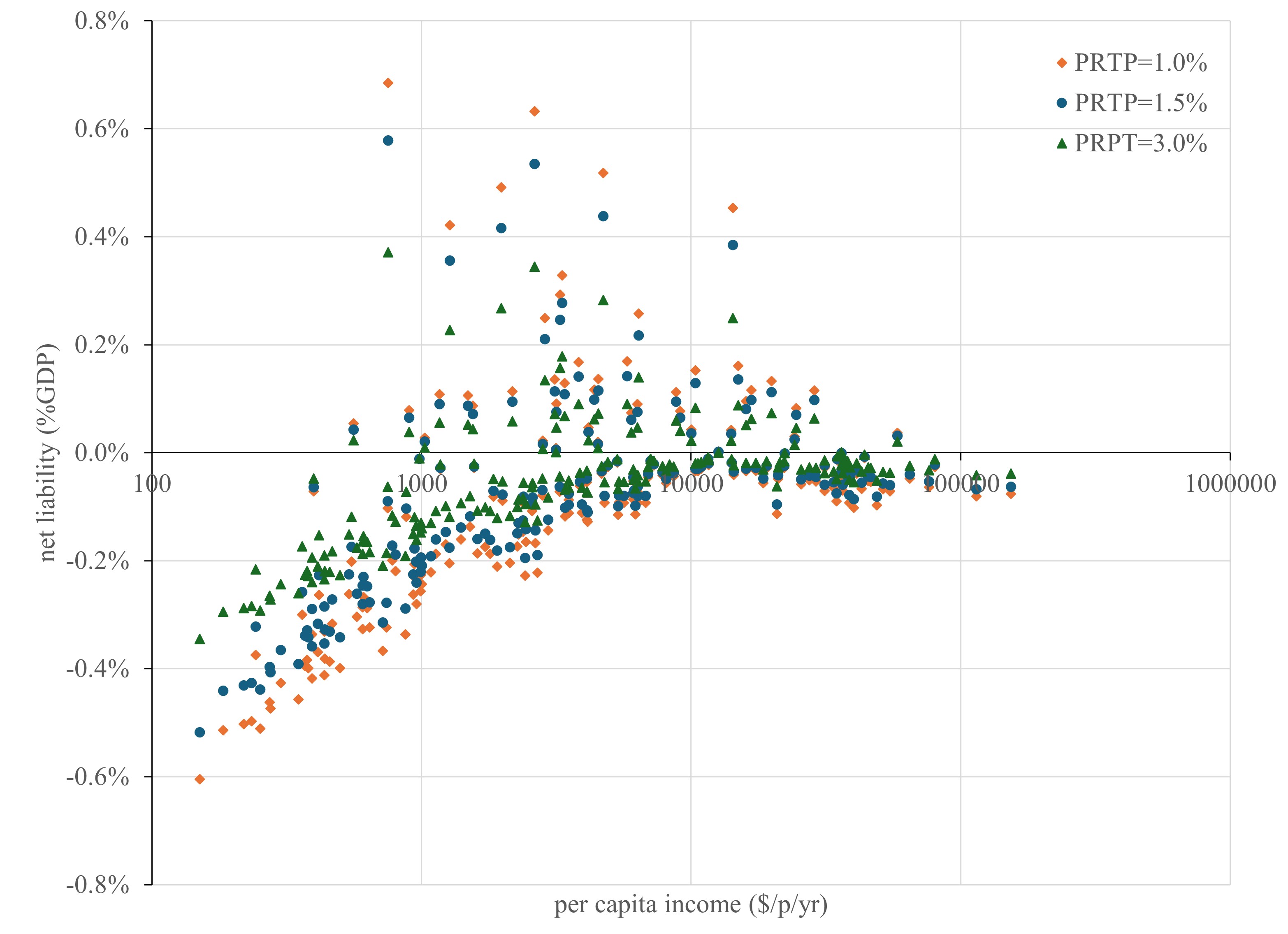}
    \includegraphics[width=0.49\linewidth]{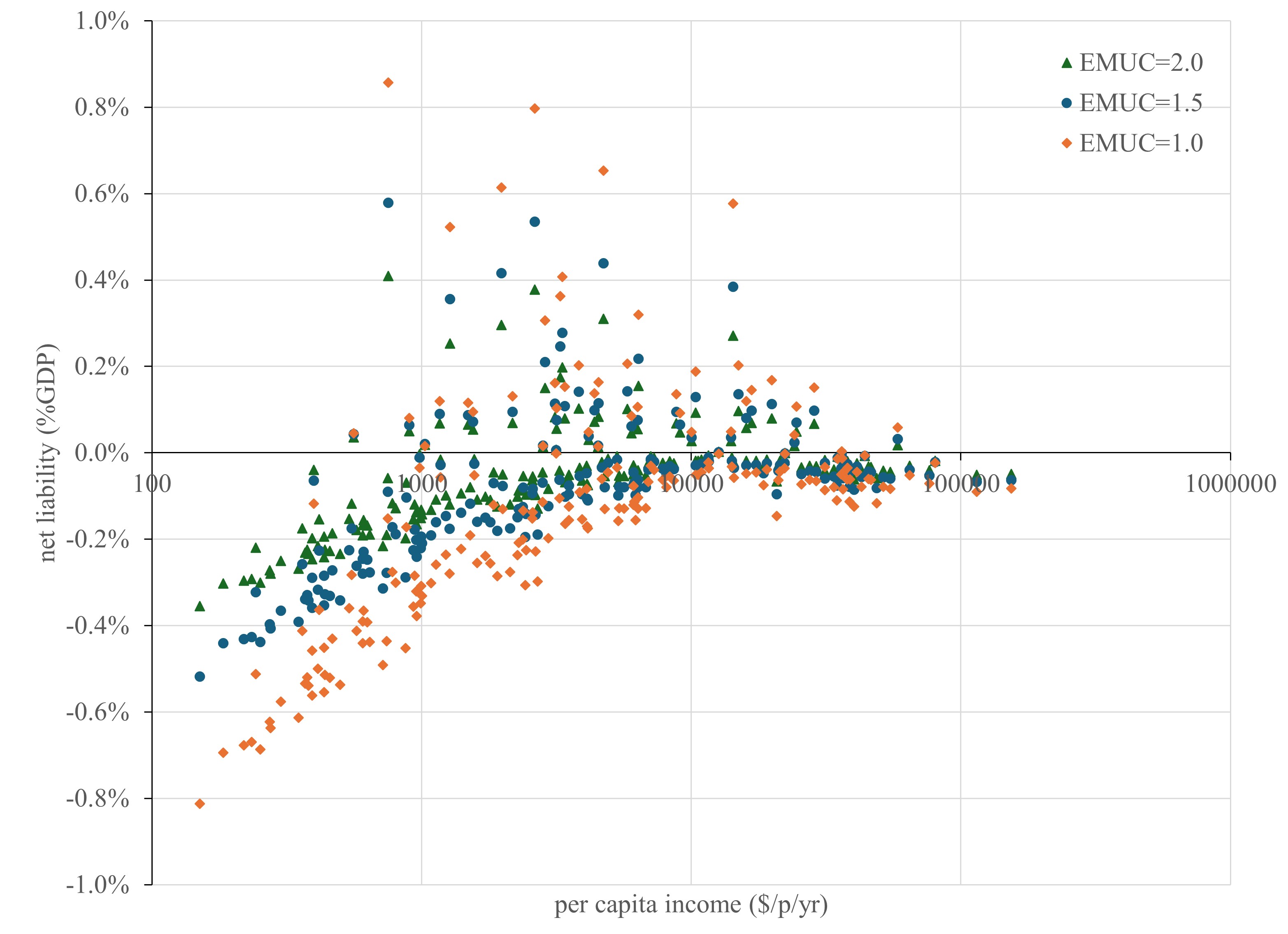}
    \caption*{\footnotesize The top left panel varies the scenario: SRES A2, RFF (default), SSP5; the top right panel compares the RFF scenario with (default) and without income convergence. The bottom left panel varies the pure rate of time preference: 1.0\%, 1.5\% (default), 3.0\%; the bottom right panel the elasticity of marginal utility of consumption: 1.0, 1.5 (default), 2.0; results are for 2010, dollars are 2005 U.S. dollars.}
\end{figure}

Figure \ref{fig:liasens} repeats Figure \ref{fig:liability}; the blue dots in both graphs are identical. The top left panel varies the scenario. There are 10 scenarios in the model. The RFF/SSP2 scenario is the default. The SRES A2 scenario has the highest absolute values of net relative liabilities. This scenario combines relatively low economic growth with relative high climate change, and thus has high impacts and high national costs of carbon. The SSP5 scenario has the lowest liabilities. This scenario combines rapid economic growth with rapid climate change. The former dominates the latter.

\begin{figure}[hp]
    \centering
    \caption{Net liability: Further sensitivity analysis.}
    \label{fig:liasens2}
    \includegraphics[width=0.49\linewidth]{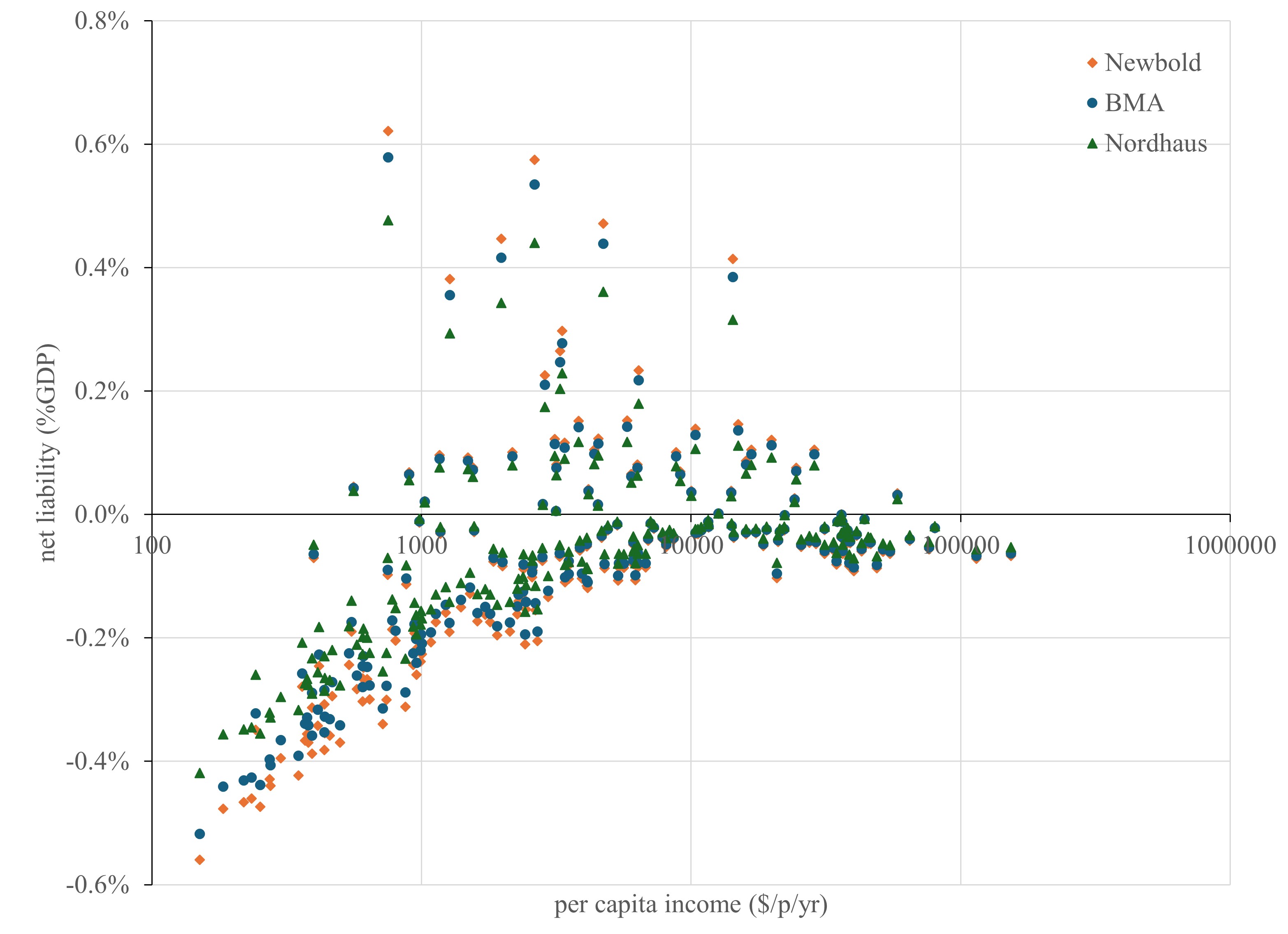}
    \includegraphics[width=0.49\linewidth]{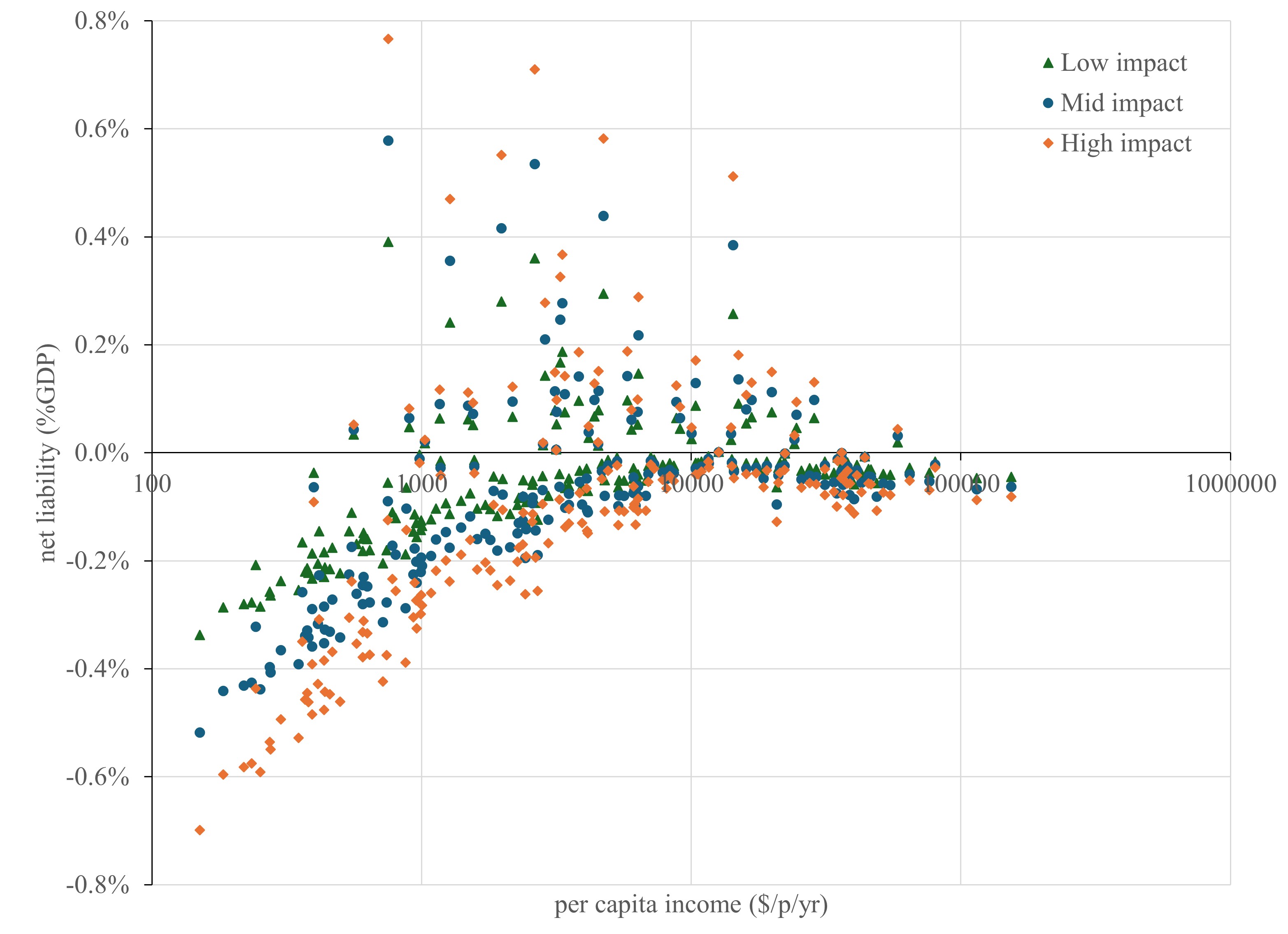}
    \includegraphics[width=0.49\linewidth]{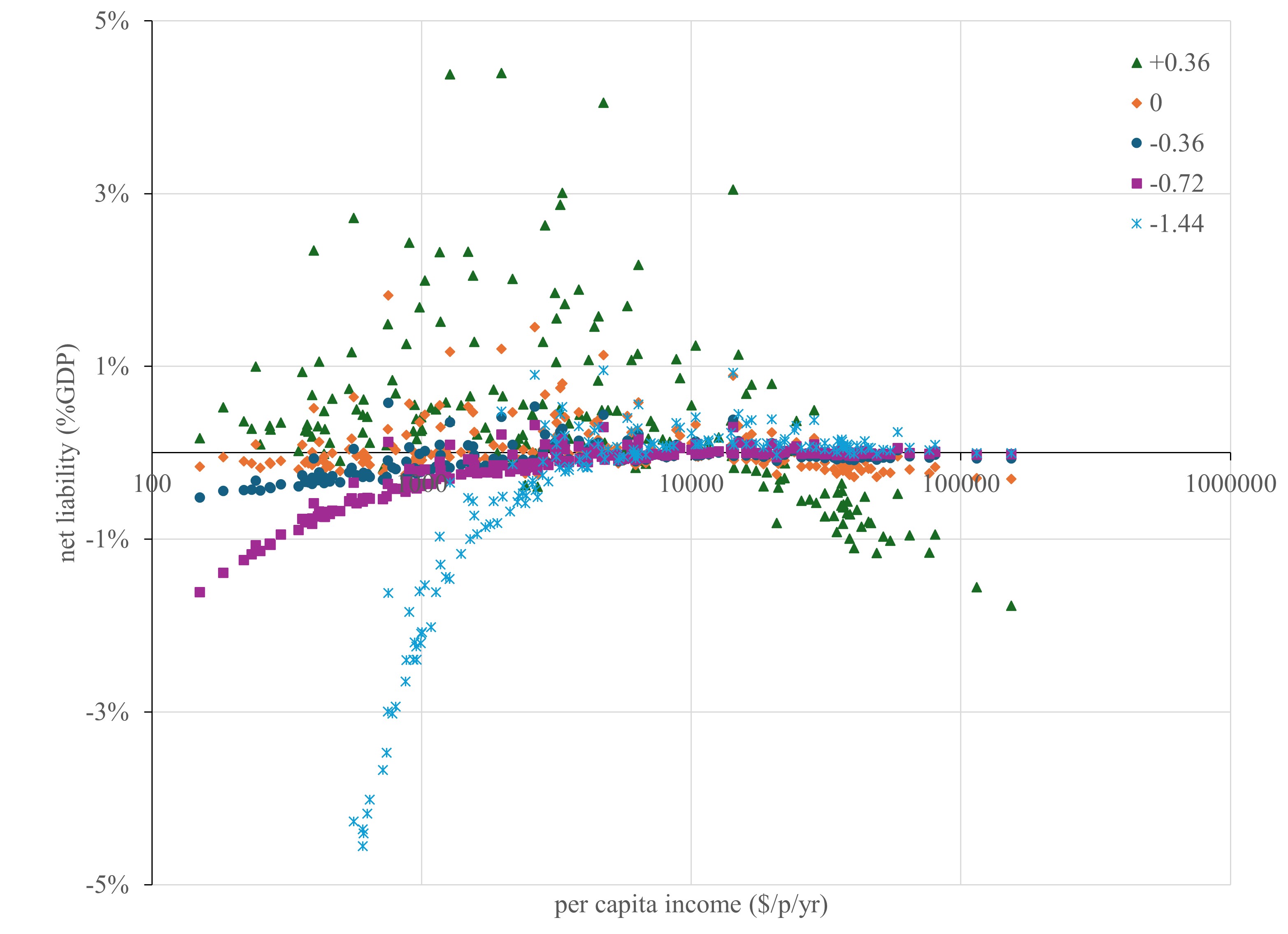}
    \includegraphics[width=0.49\linewidth]{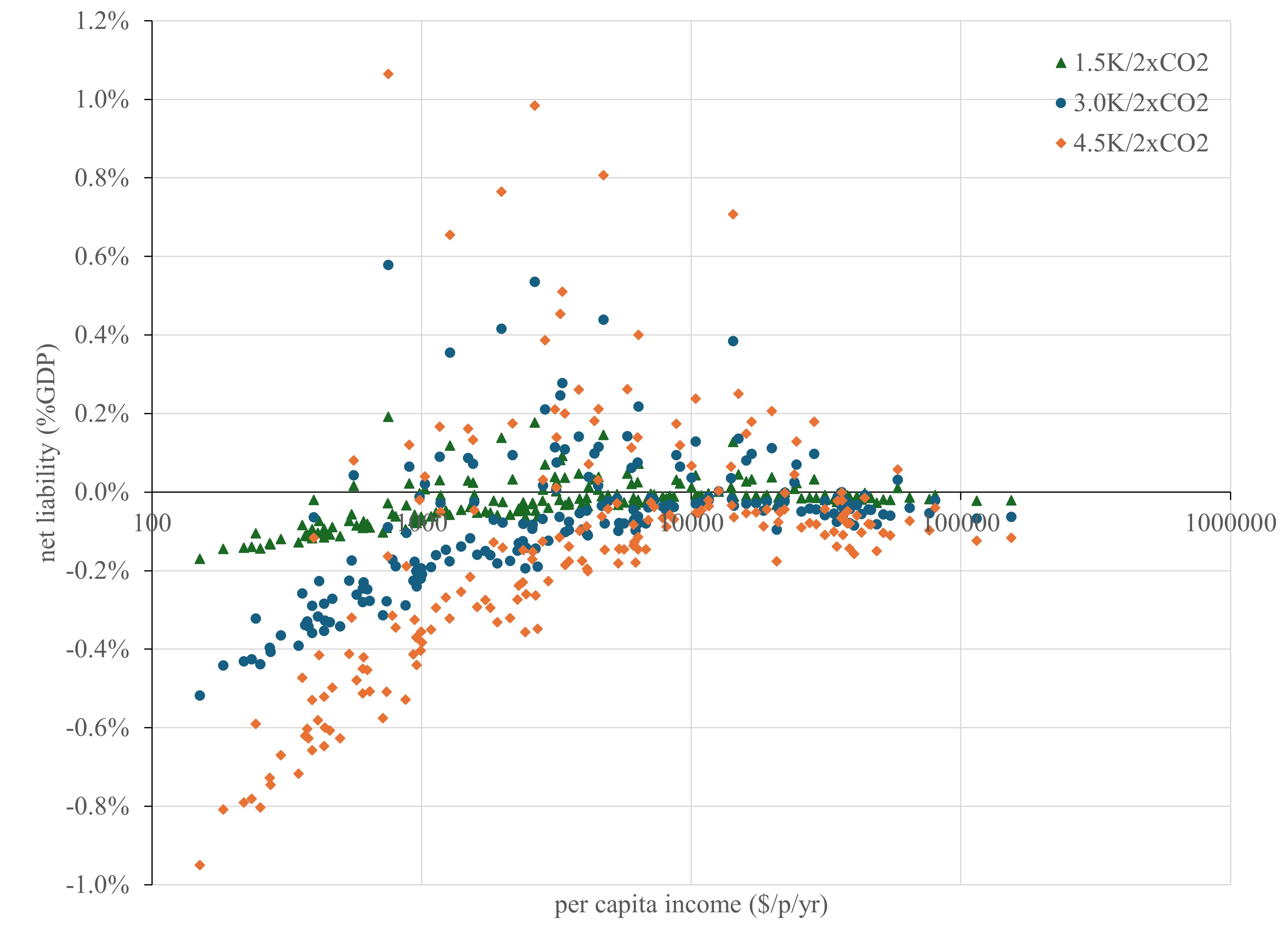}
    \caption*{\footnotesize The top left panel varies the impact function: Newbold, Bayesian model average (default), Nordhaus; the top right panel the benchmark impact: central estimates plus or minus twice the standard deviation. The bottom left panel varies the income elasticity of impact: 0.36, 0, -0.36 (default), -0.72, -1.44; the bottom right panel the climate sensitivity: 1.5\celsius, 3.0\celsius{} (default), 4.5\celsius{} per doubling of atmospheric carbon dioxide; results are for 2010, dollars are 2005 U.S. dollars.}
\end{figure}

This is highlighted in the bottom left panel of Figure \ref{fig:liasens2}, which raises the income elasticity of vulnerability to climate change from the default -0.36 to 0 and +0.36. This increases the impact of climate change and social cost of carbon in rich economies. Energy-intensive middle income countries would have to pay much more in compensation. We then double the income elasticity to -0.72. Payments to rich countries shrink; middle-income countries continue to compensate poor countries. Doubling the income elasticity again, to -1.44, rich countries pay too and compensation gets large, over a third of (pre-transfer) income in Burundi.

The top right panel of Figure \ref{fig:liasens} shows the impact of the assumption that incomes converge. If all countries grow at the same rate, the social cost of carbon is larger (smaller) in countries with an income below (above) \$25,000 per person per year; see Figure \ref{fig:convergence}. Slower growth increases vulnerability. However, since \emph{global} growth is the same between the two scenarios, richer countries grow faster if there is no convergence and their social cost of carbon is consequently lower. The effect on net liability is shown in Figure \ref{fig:liasens}: Poor countries get more compensation as their impacts are higher; middle-income countries pay less, as rich countries get less compensation.

The bottom left panel of Figure \ref{fig:liasens} varies the pure rate of time preference. The social cost of carbon goes up (down) if the pure rate of time preference goes down (up). The bottom right panel shows sensitivity to the elasticity of the marginal utility of consumption. A lower (higher) value implies a higher (lower) social cost of carbon. Net liabilities follow.

The bottom right panel of Figure \ref{fig:liasens2}, too, is easy to interpret. The social cost of carbon is higher (lower) if the climate sensitivity is higher (lower) than the default value of 3\celsius{} equilibrium warming for a doubling of the atmospheric concentration of carbon dioxide. Net liabilities follow.

The top left panel of Figure \ref{fig:liasens2} replaces the default Bayesian model average of the 13 impact function with two of its components. The estimated Nordhaus function has the lowest social cost of carbon and hence the lowest net liabilities (whether positive or negative) while the estimated Newbold function has the highest social cost of carbon and the highest liabilities. The top right panel shows the Bayesian model average plus or minus twice its standard deviation. Again, higher (lower) impacts increase (decrease) the absolute liabilities.

In all cases, the pattern of net liabilities is the same. This is because emissions partly determine net liability; see Equation (\ref{eq:netliability}. Emissions are past emissions, and therefore fixed. Carbon-intensive countries face greater liabilities. The national cost of carbon is the other determinant of net liabilities. In all but one case, larger economies see a higher national cost of carbon because there is more that can be damaged by climate change; while poorer economies see a higher national cost of carbon because they are more vulnerable. Magnitudes differ between parameter choices and scenarios but the pattern stays the same. The one exception is a positive income elasticity, which makes larger and richer economies see a higher national cost of carbon, which reinforces the pattern observed in Figure \ref{fig:liability}.

\subsection{Liability over time}
Figure \ref{fig:growth} shows the growth rate of the national cost of carbon over the period 2015-2055. Overall, the social cost of carbon grows by some 2\% per year in the poorest countries; this falls steadily to about 0.5\% in the richest countries. Countries in the Middle East are the exception: Impacts are large because it is hot already and economic activity is often at or near the coast.

\begin{figure}[hp]
    \centering
    \caption{Net liability in 2015 and 2055 plotted against per capita income in 2015.}
    \label{fig:2055}
    \includegraphics[width=1.0\linewidth]{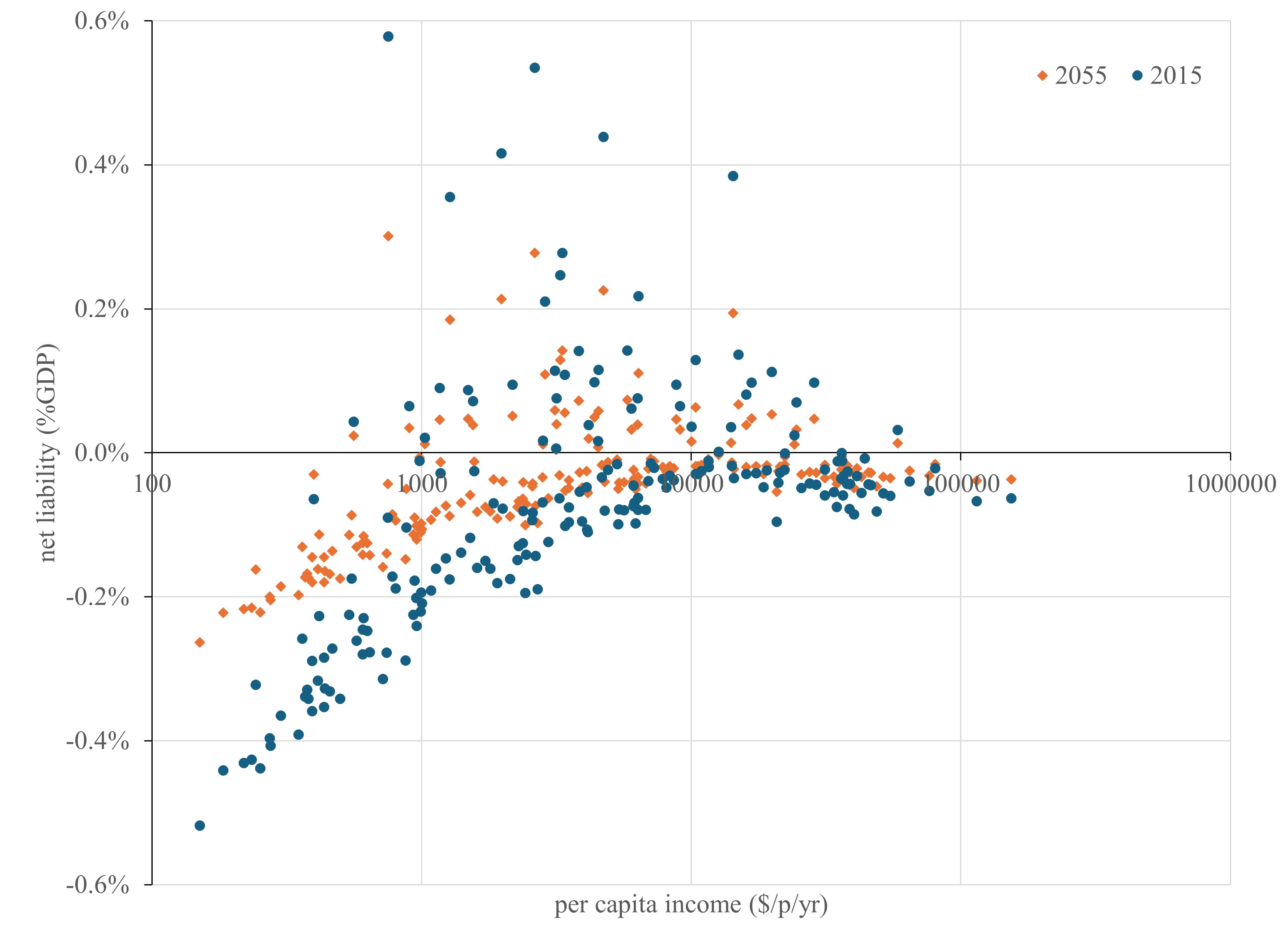}
    \caption*{\footnotesize Scenario: RFF; PRTP: 1.5\%; EMUC: 1.5; impact: Bayesian model average; income elasticity of impact: -0.36; results are for 2010, dollars are 2005 U.S. dollars.}
\end{figure}

Figure \ref{fig:2055} shows net liability in 2015 (default) and 2055. Because the social cost of carbon grows more slowly than the economy in most countries, and because carbon emissions gently decline in the RFF/SSP2 scenario, net liabilities, relative to GDP, get closer to zero.

\subsection{Historical debt}
Above, we consider liability for future damages from current emissions. This is what an accountant would count as a provisional loss. The debate on loss and damage, or rather compensation for loss and damage, is focused on the impact of past emissions \citep{James2014, Vanhala2016}.

In order to estimate historical debt, we compute the marginal impact of emissions from 1960 to 2015. Data from before 1960 are problematic. We then aggregate these impacts to 2015, using the Ramsey rule for the interest rate as above. Specifically
\begin{equation}
    D_{c,2015} = \sum_{t=1960}^{2015} \sum_{s=t}^{2015} \frac{\partial I_{c,s}}{\partial M_{t}} (1+r_c)^{2015-s}
\end{equation}
where $D_c$ is the gross debt of country $c$, $I$ are impacts, $M$ are emissions, and $r$ is the interest rate. Note that past impacts are not discounted. Instead, interest is charged on unpaid debt.

Figure \ref{fig:margdebt} plots the global historical debt
\begin{equation}
    D_t = \sum_c \sum_{s=t}^{2015} \frac{\partial I_{c,s}}{\partial M_{t}} (1+r_c)^{2015-s}
\end{equation}
against $t$. $D_t$ is very small for recent years\textemdash for 2015 emissions, only the damage in 2015 is counted\textemdash but grows steadily for earlier years, reaching some \$25/tC at the start of the period.

\begin{figure}[hp]
    \centering
    \caption{Net liability in 2015 for emissions 1960-2015.}
    \label{fig:debt}
    \includegraphics[width=1.0\linewidth]{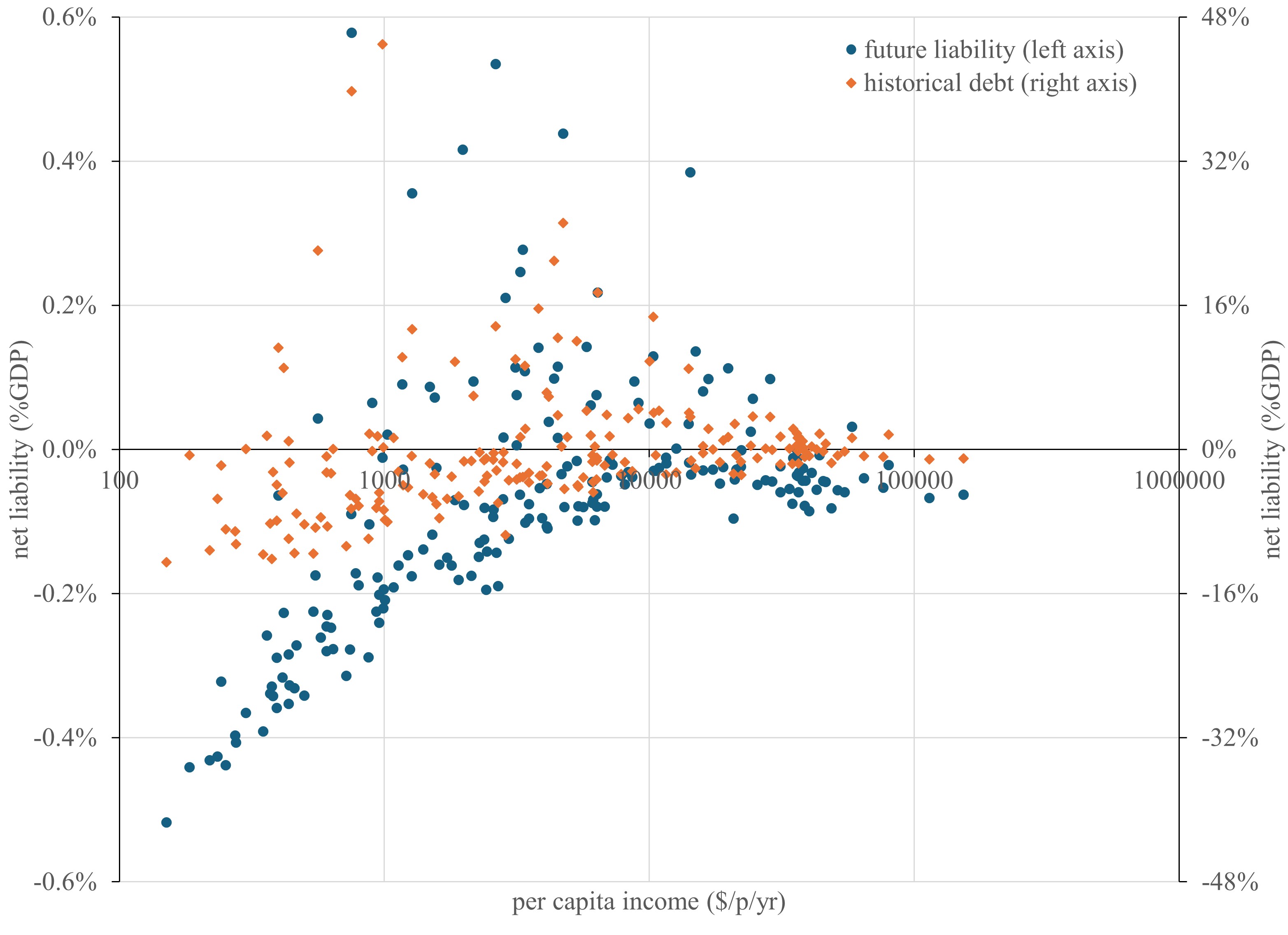}
    \caption*{\footnotesize Scenario: RFF; PRTP: 1.5\%; EMUC: 1.5; impact: Bayesian model average; income elasticity of impact: -0.36; results are for 2010, dollars are 2005 U.S. dollars.}
\end{figure}

Net historical debt follows from the analogue to Equation (\ref{eq:netliability}). Figure \ref{fig:debt} shows the results. For comparison, net liability for future damages is shown too. Magnitudes are different. This is because net liability is net liability for emissions in a \emph{single} year. Net historical debt is for emissions over 55 years. The pattern is roughly similar because the fundamentals change only slowly over time. Carbon-intensive middle-income countries have built up a considerable historical debt. Poor countries have emitted little and are particularly vulnerable to climate change. Emissions are relatively low in rich countries as are their impacts, but the latter count for a lot if accounted for in dollars.

\section{Discussion and conclusion}
\label{sc:conclude}
We present new estimates of the national cost of carbon, reflecting a recently updated meta-analysis of the total impact of climate change. The social cost of carbon is larger in countries with more people, and in countries where people are poorer. We use these estimates to assess the net liability for climate impacts, defined as the harm done to others minus the harm suffered. The net liability is positive in middle-income countries with carbon-intensive economies; it is negative in the richest and poorest countries. Both rich and poor suffer large absolute damages from climate change\textemdash the latter because they are vulnerable, the former because their damages count for a lot\textemdash while contributing relatively little to current emissions. The compensation that could be demanded is relatively large (small) in poor (rich) countries. Historical debt shows the same pattern as liability for future impacts, but magnitudes are larger since debt has accumulated over time.

Liability for the impact of climate change is an academic question. This may change if current attempts to hold emitters responsible succeed in court. That would also clarify who is responsible\textemdash producers, consumers, regulators\textemdash for what\textemdash damages, net damages, material damages, distress\textemdash since when\textemdash 1827, 1896, 1985, 1995\textemdash and for potential future harm or actual past damages. In the current paper, we assume that liability is for net future damages from current emissions, and assign this to countries as a whole.

The perhaps surprising conclusion that rich countries should be compensated rests on two assumptions: First, in the main analysis, we assume that current emissions matter rather than historical emissions. Past climate change caused less damage than future climate change will; see Figure \ref{fig:impact}. Rich countries are largely responsible for cumulative emissions, but less so for current concentrations, and even less so for future climate change.

The second assumption is that the social cost of carbon is a Lindahl price \citep{Kelleher2025}: The global social cost of carbon is the sum of the national social costs of carbon, expressed in dollars. As compensation would be paid in money, it seems logical to express the social cost of carbon in money too. However, an equity-weighted social cost of carbon would not only be much larger, but would also shift its distribution decidedly towards poorer countries \citep{Fankhauser1997}. That said, there is a Pareto-superior deal in which the rich compensate the poor for accepting the Lindahl price \citep{Schelling1986}.

More research is needed into these matters. The other papers in this special issue explore another key dimension, the relative and absolute size of the national social cost of carbon. As it stands, there appears to be potential support for a coalition of rich and poor advocating for liability for the damages of climate change.

\section*{Acknowledgements}
Financial support by the Seoul Institute and the Korea Ministry for the Environment (RS-2023-00218794) is gratefully acknowledged. Diane Coyle and Robert Mendelsohn had excellent comments on an earlier draft. All errors and opinions are ours.

\section*{Author contributions}
MA \& RT jointly conceptualized the study. RT programmed and ran the code, and wrote the first draft. MA \& RT edited the final draft. 

\section*{Competing interest statement}
None.

\section*{Data availability statement}
The data is available with the code as indicated below.

\section*{Code availability statement}
Data and code are on \href{https://github.com/rtol/FUND-National}{GitHub}.

\newpage

\newpage \begin{table}[hp]
    \centering
    \caption{Net liability, billion US dollar per year}
    \label{tab:liability}
    \begin{tabular}{l r l r}
    \hline
\multicolumn{2}{c}{biggest winners} & \multicolumn{2}{c}{biggest losers} \\ \hline
Japan	&	-1.72	&	China	&	8.07	\\
France	&	-1.20	&	Russia	&	1.98	\\
Germany	&	-1.06	&	Iran	&	0.60	\\
United States	&	-1.05	&	South Africa	&	0.41	\\
United Kingdom	&	-1.02	&	Ukraine	&	0.38	\\
Brazil	&	-0.88	&	Saudi Arabia	&	0.35	\\
Italy	&	-0.78	&	Kazakhstan	&	0.34	\\
Spain	&	-0.58	&	India	&	0.26	\\
Mexico	&	-0.46	&	Thailand	&	0.16	\\
Nigeria	&	-0.31	&	Poland	&	0.14	\\
Switzerland	&	-0.26	&	Malaysia	&	0.14	\\
Sweden	&	-0.22	&	Uzbekistan	&	0.12	\\
Netherlands	&	-0.22	&	Venezuela	&	0.11	\\
Turkey	&	-0.21	&	Egypt	&	0.09	\\
Bangladesh	&	-0.18	&	Kuwait	&	0.08	\\
 \hline
    \end{tabular}
    \caption*{\footnotesize Dollars are 2005 U.S. dollars.}
\end{table}

\newpage \bibliography{master}

\newpage \appendix
\setcounter{page}{1}
\renewcommand{\thepage}{A\arabic{page}}
\setcounter{table}{0}
\renewcommand{\thetable}{A\arabic{table}}
\setcounter{figure}{0}
\renewcommand{\thefigure}{A\arabic{figure}}
\setcounter{equation}{0}
\renewcommand{\theequation}{A\arabic{equation}}

\begin{figure}[h!]
    \centering
    \caption{The national social cost of carbon plotted against population size in 2010.}
    \label{fig:pop}
    \includegraphics[width=1.0\linewidth]{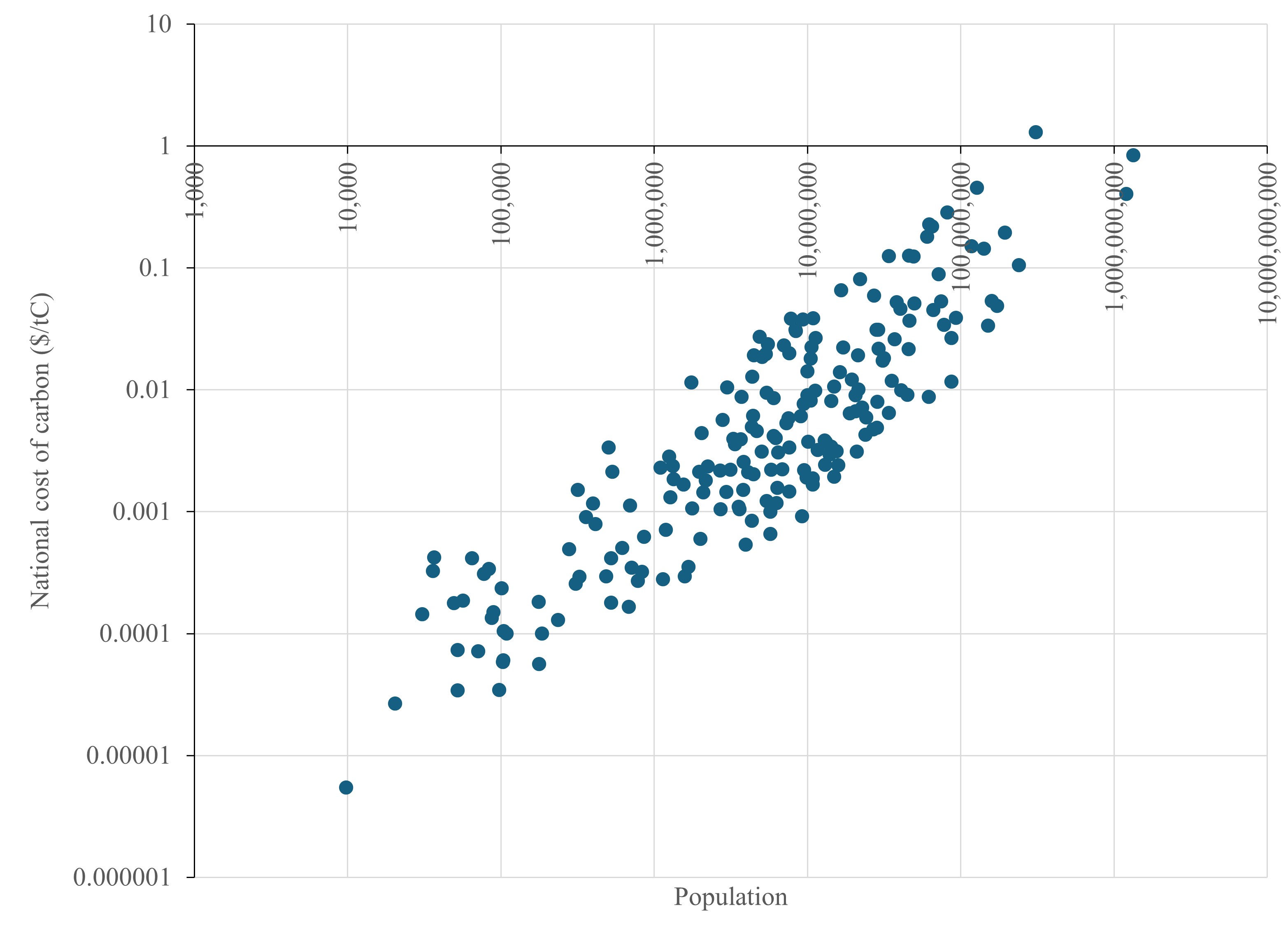}
    \caption*{\footnotesize Scenario: RFF; PRTP: 1.5\%; EIS: 1.5; impact: Bayesian model average; income elasticity of impact: -0.36; results are for 2010.}
\end{figure}

\begin{figure}[h!]
    \centering
    \caption{The national average social cost of carbon plotted against per capita income in 2010 for three alternative income elasticities of impact.}
    \label{fig:incelas}
    \includegraphics[width=1.0\linewidth]{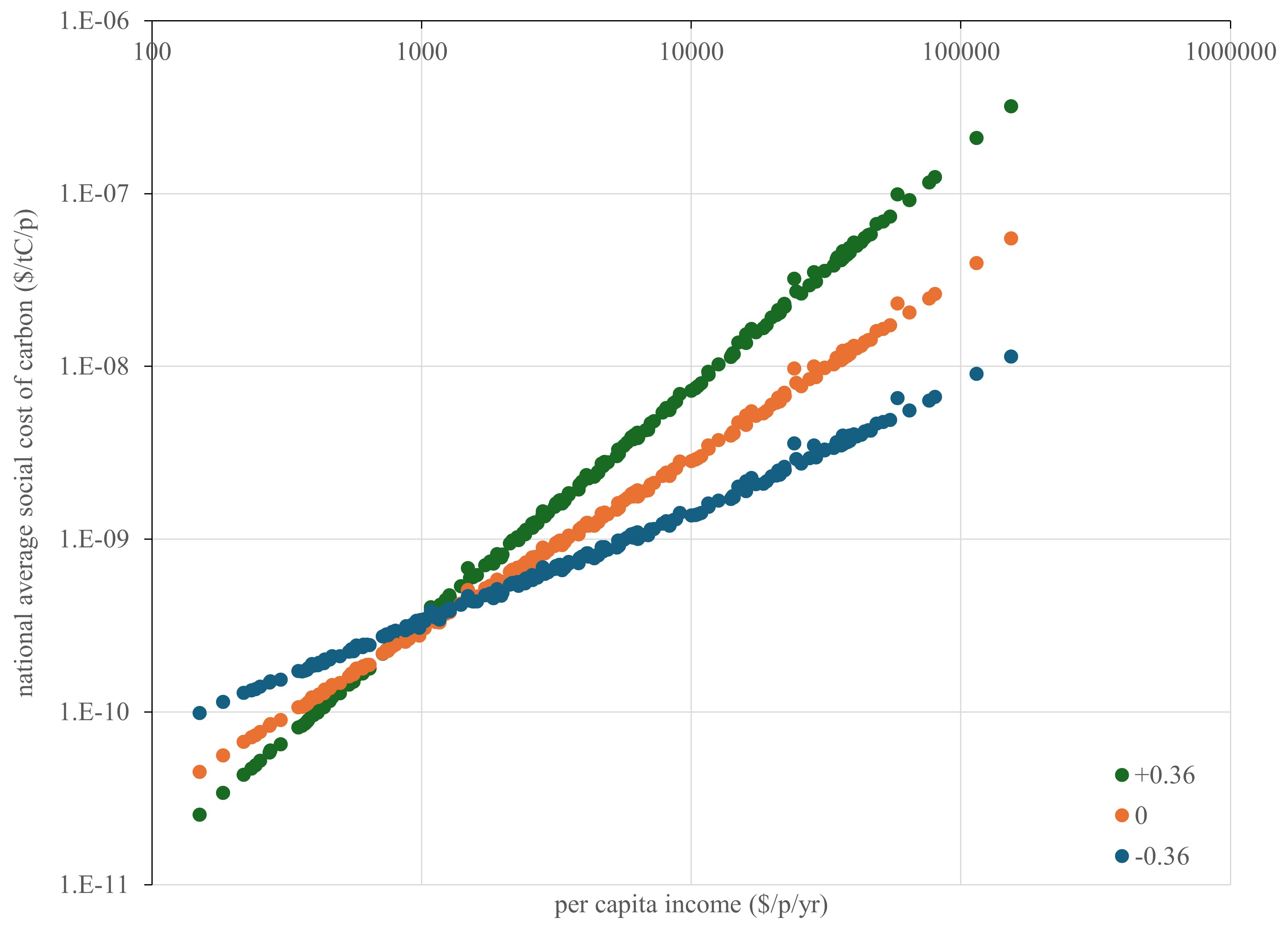}
    \caption*{\footnotesize Scenario: RFF; PRTP: 1.5\%; EIS: 1.5; impact: Bayesian model average; income elasticity of impact: -0.36, 0, +0.36; per capita income is in 2005 U.S. dollars for 2010.}
\end{figure}

\begin{figure}[h!]
    \centering
    \caption{The national social cost of carbon plotted against carbon efficiency.}
    \label{fig:carbon}
    \includegraphics[width=1.0\linewidth]{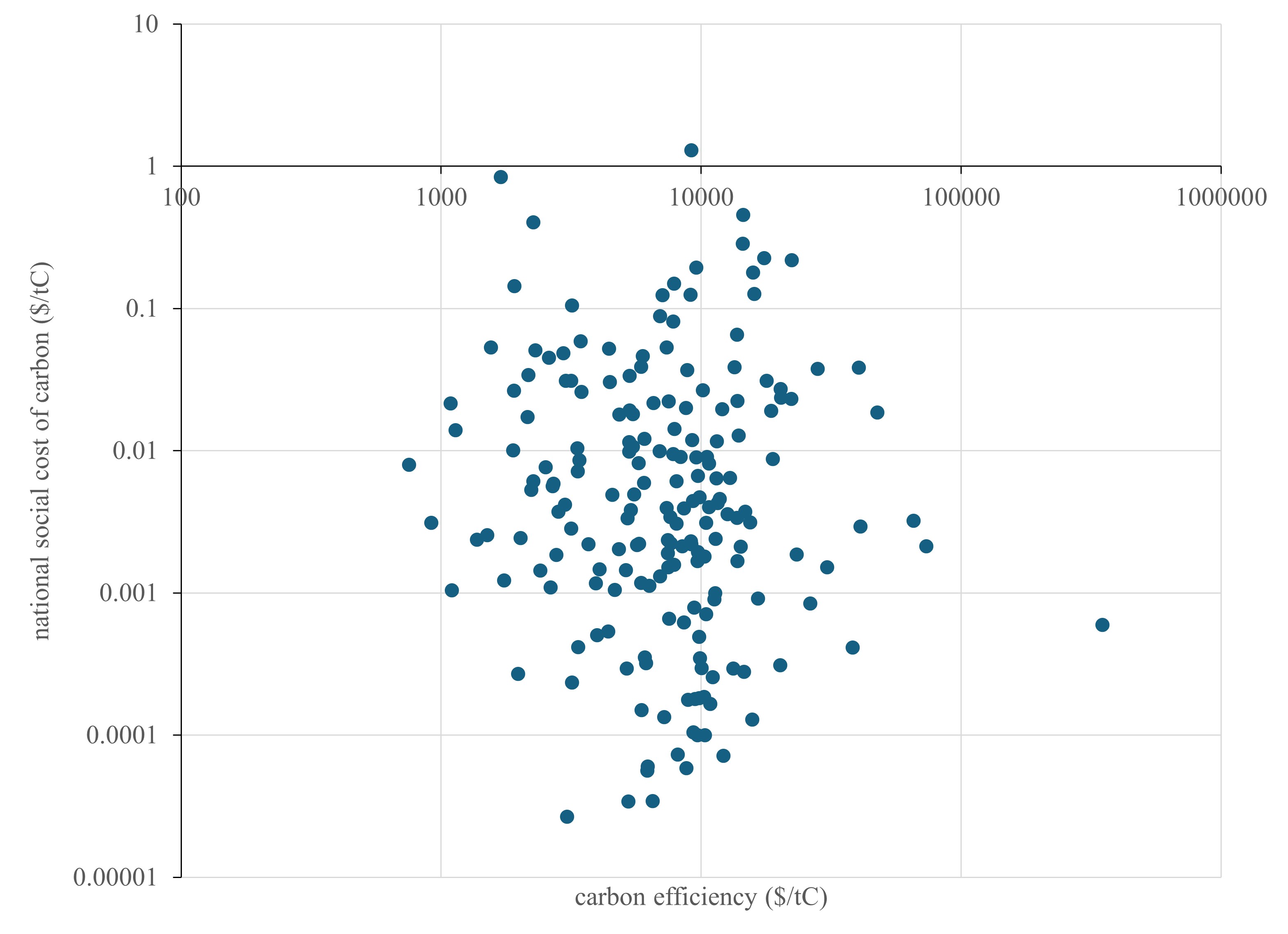}
    \caption*{\footnotesize Scenario: RFF; PRTP: 1.5\%; EIS: 1.5; impact: Bayesian model average; income elasticity of impact: -0.36; carbon efficiency is in 2005 U.S. dollars for 2010.}
\end{figure}

\begin{figure}[h!]
    \centering
    \caption{The damage imposed \emph{on} other countries plotted against per capita income.}
    \label{fig:harm}
    \includegraphics[width=1.0\linewidth]{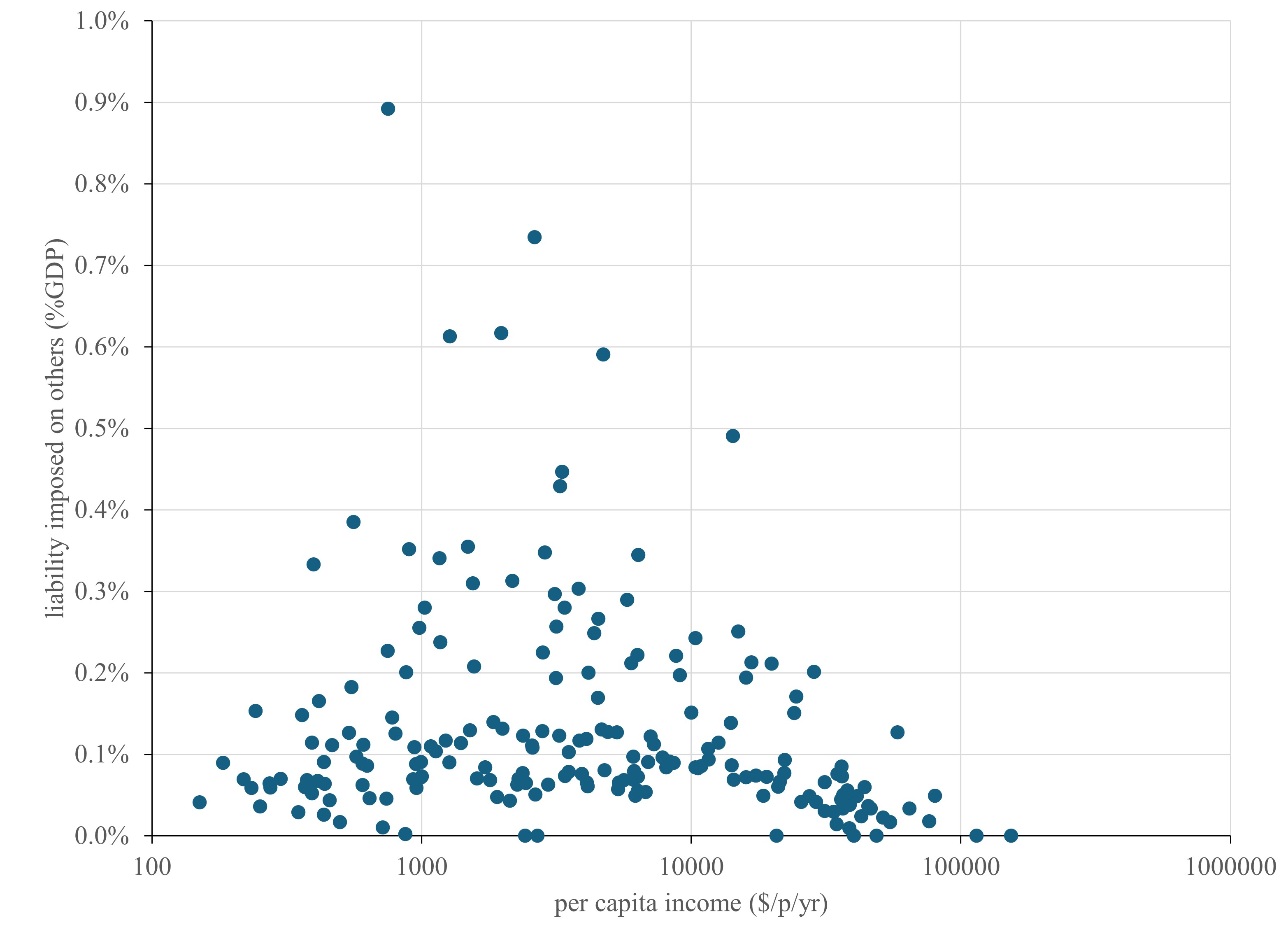}
    \caption*{\footnotesize Scenario: RFF; PRTP: 1.5\%; EIS: 1.5; impact: Bayesian model average; income elasticity of impact: -0.36; per capita income is in 2005 U.S. dollars for 2010.}
\end{figure}

\begin{figure}[h!]
    \centering
    \caption{The damage imposed \emph{by} other countries plotted against per capita income.}
    \label{fig:damage}
    \includegraphics[width=1.0\linewidth]{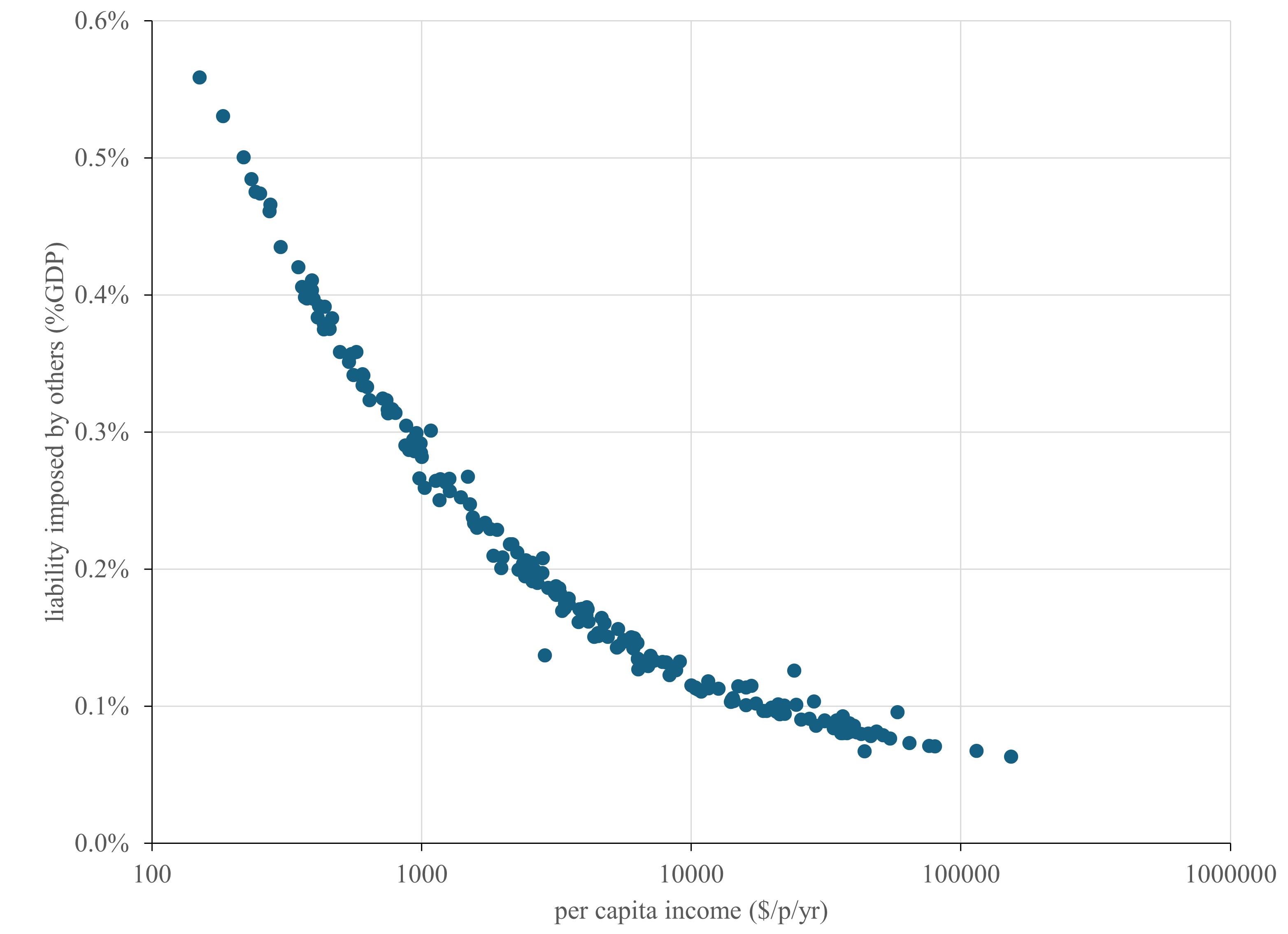}
    \caption*{\footnotesize Scenario: RFF; PRTP: 1.5\%; EIS: 1.5; impact: Bayesian model average; income elasticity of impact: -0.36; per capita income is in 2005 U.S. dollars for 2010.}
\end{figure}

\begin{figure}[h!]
    \centering
    \caption{The national share in global emissions and social cost of carbon plotted against per capita income.}
    \label{fig:share}
    \includegraphics[width=1.0\linewidth]{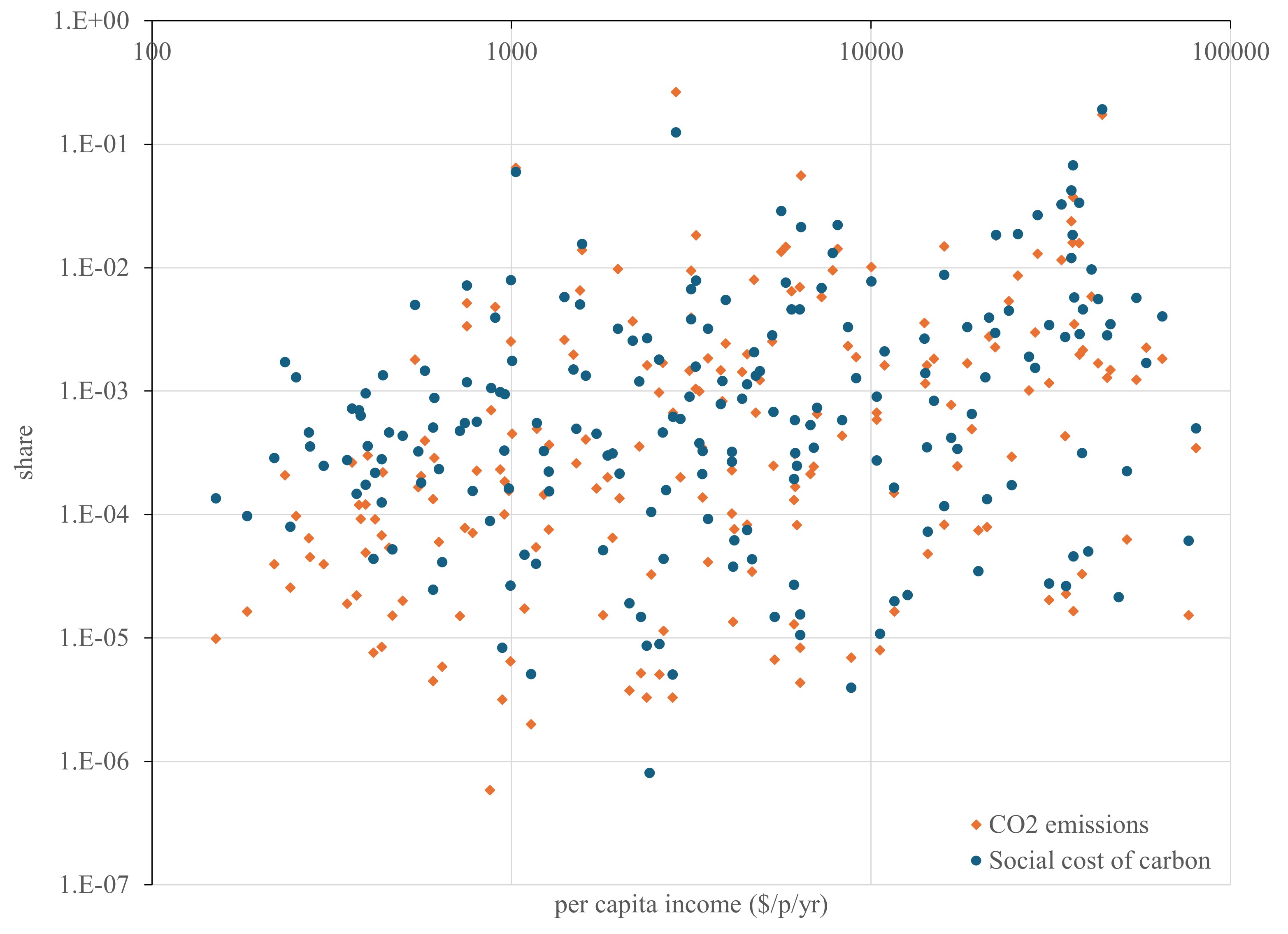}
    \caption*{\footnotesize Scenario: RFF; PRTP: 1.5\%; EIS: 1.5; impact: Bayesian model average; income elasticity of impact: -0.36; per capita income is in 2005 U.S. dollars for 2010.}
\end{figure}

\begin{figure}[h!]
    \centering
    \caption{The difference in the national social cost of carbon due to the assumption of income convergence, plotted against per capita income.}
    \label{fig:convergence}
    \includegraphics[width=1.0\linewidth]{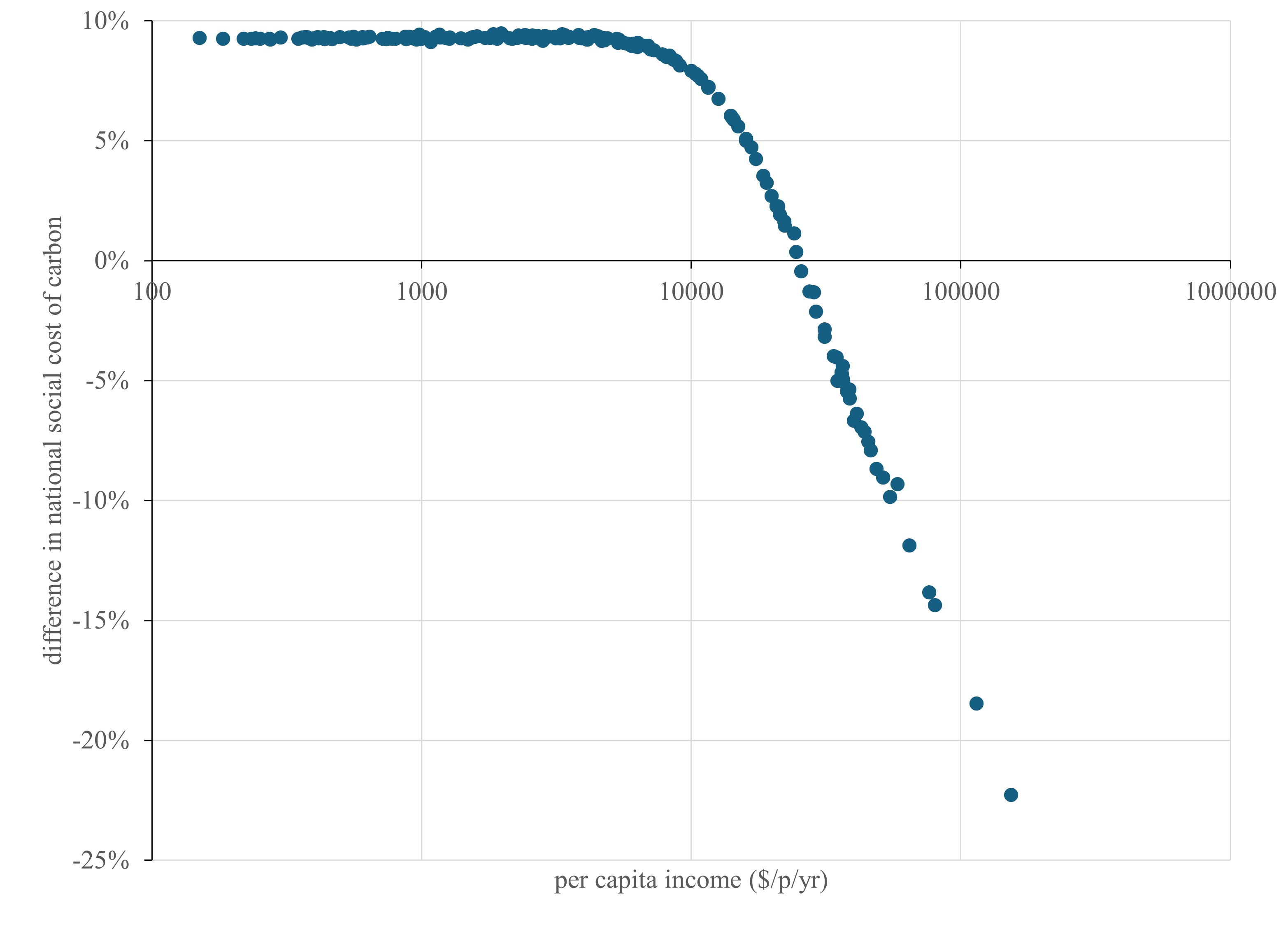}
    \caption*{\footnotesize Scenario: RFF; PRTP: 1.5\%; EIS: 1.5; impact: Bayesian model average; income elasticity of impact: -0.36; per capita income is in 2005 U.S. dollars for 2010.}
\end{figure}

\begin{figure}[h!]
    \centering
    \caption{The growth rate of the social cost of carbon over the period 2015-2055 plotted against per capita income.}
    \label{fig:growth}
    \includegraphics[width=1.0\linewidth]{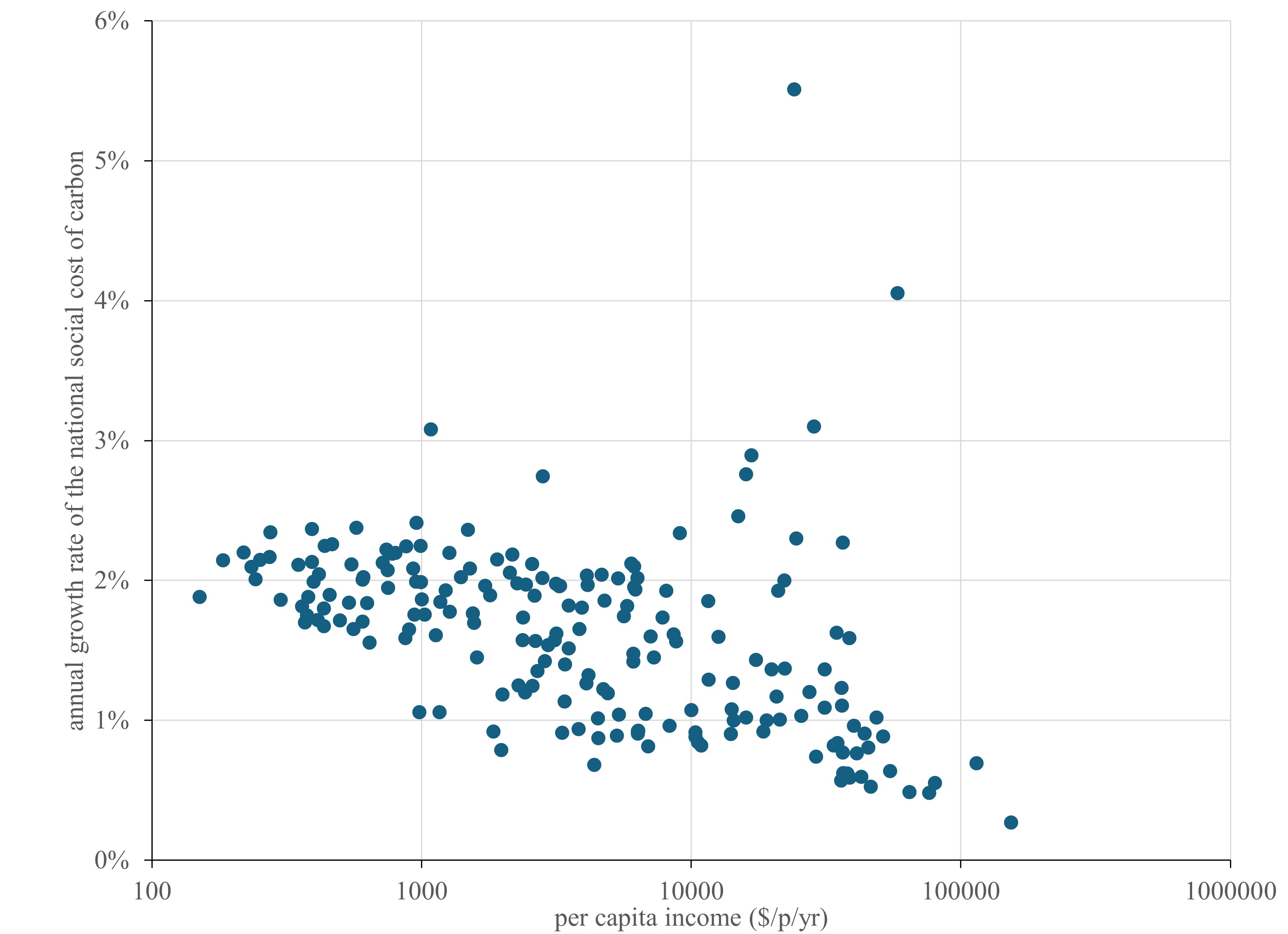}
    \caption*{\footnotesize Scenario: RFF; PRTP: 1.5\%; EIS: 1.5; impact: Bayesian model average; income elasticity of impact: -0.36; per capita income is in 2005 U.S. dollars for 2010.}
\end{figure}

\begin{figure}[h!]
    \centering
    \caption{The marginal historical debt incurred by year of emission.}
    \label{fig:margdebt}
    \includegraphics[width=1.0\linewidth]{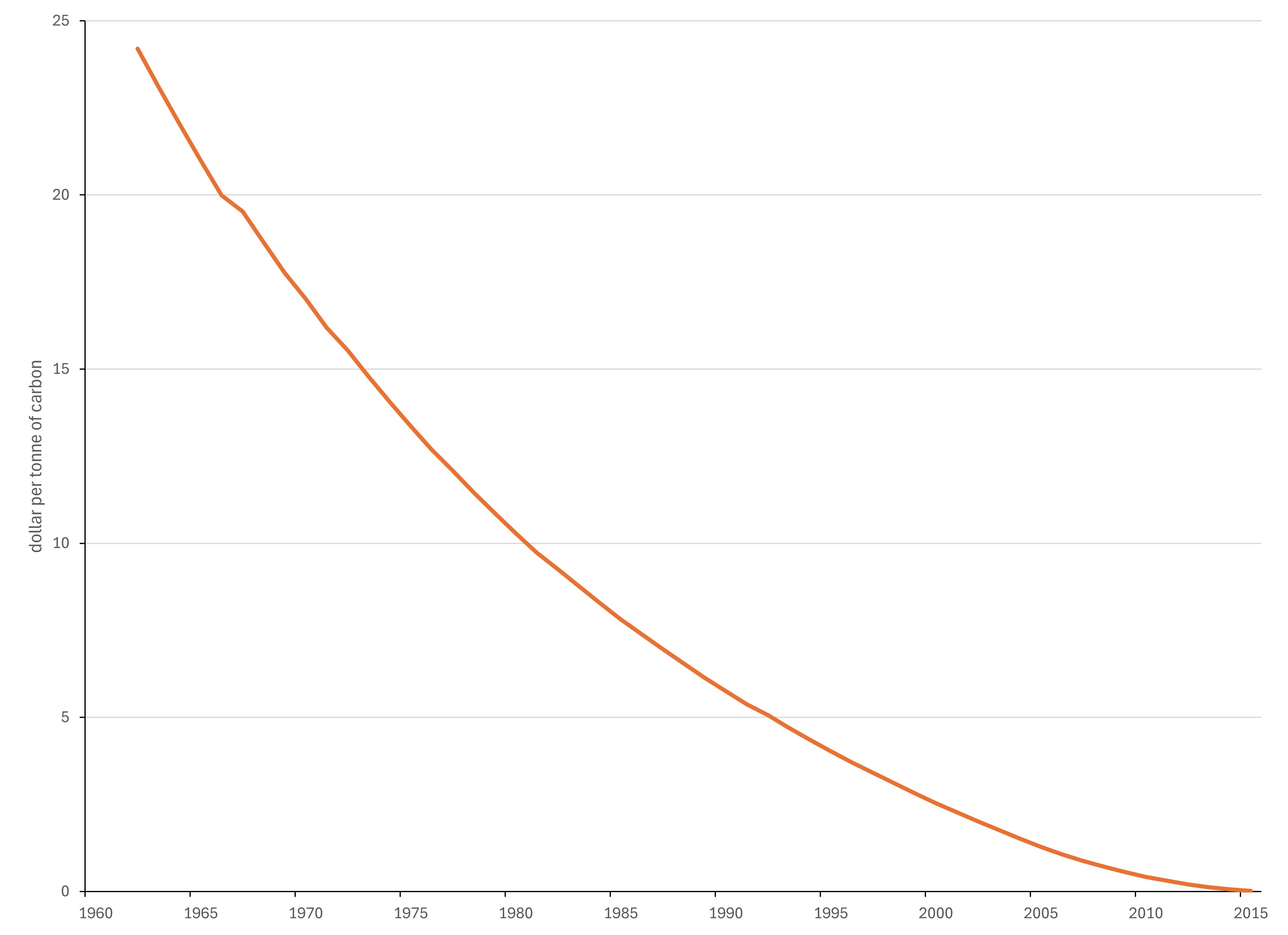}
    \caption*{\footnotesize Scenario: RFF; PRTP: 1.5\%; EIS: 1.5; impact: Bayesian model average; income elasticity of impact: -0.36; dollars are 2005 U.S. dollars.}
\end{figure}

\begin{table}[h!p]
    \centering \footnotesize
    \caption{Impact functions}
    \label{tab:function}
\begin{tabular}{|l|c|c|l|r|} \hline
    name & function  & likelihood (\%) & proponent & scc\\ \hline
Parabolic & $-0.45T -0.082T^2 $ &	19.65  & after \citet{Tol2009JEP} & 12\\
Threshold & \thead{$0 \text{ if } T \leq -0.21$ \\ $-0.49 (T+0.21)^{1.3} \text{ if }  T > -0.21$} & 18.01 & after \citet{Newbold2013} & 15\\
Piecewise linear & \thead{$0.46T \text{ if } T \leq 0.74$ \\ $0.46 \cdot 0.74 - 0.90 T  \text{ if }  T > 0.90$} & 14.31  & after \citet{Tol2018REEP} & 14\\
Quadratic & $-0.17 T^2$ & 13.15 & after \citet{Nordhaus1992} & 10\\
Hazard & $-0.19 T^2 + 1.10 \cdot 10^{-5} T^6$ &	11.12 & after \citet{Weitzman2012} & 10\\
Hazard & $-0.18 T^2+ 1.55 \cdot 10^{-6} T^7$ &	11.02  & after \citet{Weitzman2012} & 10\\
Linear & $-0.79T$ & 10.68  & after \citet{Hope2006c} & 13\\
Cubic & $-0.028 T^2$ & 1.60 & after \citet{Peck1995} & 11\\
Quartic & $-0.0039 T^2$ & 0.37 & after \citet{Azar2003} & 2\\
Exponential & $0.0078 - 0.0078e^T$ & 0.09 & after \citet{Ploeg2014EER} & 1 \\ Quadratic & $-0.35 T^2$ & $7.1 \cdot 10^{-4}$ & \citet{Barrage2024} & 19\\
Quadratic & $-0.86 T^2$ & $5.8 \cdot 10^{-20}$ & \citet{Howard2017} & 47\\
Hazard & $-0.24 T^2 - 0.00051 T^{6.754}$ &	$4.3 \cdot 10^{-47}$ & \citet{Weitzman2012} & 17\\
\hline
\end{tabular}
\caption*{\scriptsize The social cost of carbon, in 2005 U.S. dollar per metric tonne of carbon, is for emissions in the year 2015, the RFF/SSP2 scenario, a pure rate of time preference of 1.0\%, and a rate of risk aversion of 1.5.}  
\end{table}

\tiny
\begin{center}
\begin{longtable}{lrrrrrrrrrrrr}
\caption{The national cost of carbon for various scenarios} \label{tab:sccscen} \\

\hline \multicolumn{1}{l}{country} & \multicolumn{1}{c}{RFF} & \multicolumn{1}{c}{A1} & \multicolumn{1}{c}{A2} & \multicolumn{1}{c}{B1} & \multicolumn{1}{c}{B2} & \multicolumn{1}{c}{SSP1} & \multicolumn{1}{c}{SSP2} & \multicolumn{1}{c}{SSP3} & \multicolumn{1}{c}{SSP4} & \multicolumn{1}{c}{SSP5}& \multicolumn{1}{c}{no conv} & \multicolumn{1}{c}{2055}\\ \hline 
\endfirsthead

\multicolumn{13}{c}%
{{\bfseries \tablename\ \thetable{} -- continued from previous page}} \\
\hline \multicolumn{1}{l}{country} & \multicolumn{1}{c}{RFF} & \multicolumn{1}{c}{A1} & \multicolumn{1}{c}{A2} & \multicolumn{1}{c}{B1} & \multicolumn{1}{c}{B2} & \multicolumn{1}{c}{SSP1} & \multicolumn{1}{c}{SSP2} & \multicolumn{1}{c}{SSP3} & \multicolumn{1}{c}{SSP4} & \multicolumn{1}{c}{SSP5}& \multicolumn{1}{c}{no conv} & \multicolumn{1}{c}{2055}\\ \hline 
\endhead

\hline \multicolumn{13}{r}{{continued on next page}} \\ \hline
\endfoot

\hline \multicolumn{13}{l}{The left-most column has the default RFF scenario. To its right are the SRES scenarios and the SSP scenarios. The second-most column to the right} \\ \multicolumn{13}{l}{has the RFF scenario without income convergence. The right-most column shows the social cost of carbon for 2055.} \\ \hline
\endlastfoot

Afghanistan	&	0.00487	&	0.00324	&	0.00575	&	0.00383	&	0.00462	&	0.00330	&	0.00395	&	0.00533	&	0.00414	&	0.00285	&	0.00532	&	0.00999	\\
Albania	&	0.00220	&	0.00147	&	0.00260	&	0.00173	&	0.00209	&	0.00149	&	0.00178	&	0.00241	&	0.00187	&	0.00130	&	0.00240	&	0.00383	\\
Algeria	&	0.02580	&	0.01728	&	0.03044	&	0.02032	&	0.02449	&	0.01756	&	0.02097	&	0.02823	&	0.02194	&	0.01525	&	0.02819	&	0.05641	\\
Andorra	&	0.00031	&	0.00021	&	0.00036	&	0.00024	&	0.00029	&	0.00022	&	0.00026	&	0.00035	&	0.00027	&	0.00018	&	0.00030	&	0.00076	\\
Angola	&	0.01209	&	0.00809	&	0.01425	&	0.00951	&	0.01147	&	0.00822	&	0.00982	&	0.01322	&	0.01028	&	0.00714	&	0.01320	&	0.02793	\\
Antigua	&	0.00013	&	0.00009	&	0.00016	&	0.00011	&	0.00013	&	0.00009	&	0.00011	&	0.00015	&	0.00012	&	0.00008	&	0.00014	&	0.00022	\\
Argentina	&	0.04609	&	0.03104	&	0.05427	&	0.03646	&	0.04381	&	0.03172	&	0.03767	&	0.05022	&	0.03941	&	0.02753	&	0.05013	&	0.08190	\\
Armenia	&	0.00145	&	0.00096	&	0.00171	&	0.00113	&	0.00137	&	0.00098	&	0.00117	&	0.00158	&	0.00123	&	0.00084	&	0.00158	&	0.00231	\\
\end{longtable}

\end{center}
\newpage
\tiny
\begin{center}
\begin{longtable}{lrrrrrrrrrrrrrr}
\caption{The national cost of carbon for various parameter choices} \label{tab:sccsens} \\

\hline \multicolumn{1}{l}{country} & \multicolumn{1}{c}{0.36} & \multicolumn{1}{c}{0} & \multicolumn{1}{c}{-0.72} & \multicolumn{1}{c}{-1.44} & \multicolumn{1}{c}{Nord} & \multicolumn{1}{c}{Newb} & \multicolumn{1}{c}{-2SD} & \multicolumn{1}{c}{+2SD} & \multicolumn{1}{c}{1\%} & \multicolumn{1}{c}{3\%} & \multicolumn{1}{c}{1} & \multicolumn{1}{c}{2} & \multicolumn{1}{c}{1.5}& \multicolumn{1}{c}{4.5} \\ \hline 
\endfirsthead

\multicolumn{15}{c}%
{{\bfseries \tablename\ \thetable{} -- continued from previous page}} \\
\hline \multicolumn{1}{l}{country} & \multicolumn{1}{c}{0.36} & \multicolumn{1}{c}{0} & \multicolumn{1}{c}{-0.72} & \multicolumn{1}{c}{-1.44} & \multicolumn{1}{c}{Nord} & \multicolumn{1}{c}{Newb} & \multicolumn{1}{c}{-2SD} & \multicolumn{1}{c}{+2SD} & \multicolumn{1}{c}{1\%} & \multicolumn{1}{c}{1} & \multicolumn{1}{c}{2} & \multicolumn{1}{c}{3\%} & \multicolumn{1}{c}{1.5}& \multicolumn{1}{c}{4.5} \\ \hline  
\endhead

\hline \multicolumn{15}{r}{continued on next page} \\ \hline
\endfoot

\hline \multicolumn{15}{l}{The income elasticity of impacts is varied in columns 2-5; the impact function in 6-7; benchmark impact in 8-9; the pure rate of time preference in 10-11;} \\ \multicolumn{15}{l}{the elasticity of the marginal utility of consumption in 12-13; and the climate sensitivity in 14-15. The default (RFF) is in Table \ref{tab:sccscen}.}\\
\hline
\endlastfoot

Afghanistan	&	0.0024	&	0.0030	&	0.0105	&	0.1159	&	0.0039	&	0.0053	&	0.0032	&	0.0066	&	0.0057	&	0.0032	&	0.0076	&	0.0033	&	0.0016	&	0.0089	\\
Albania	&	0.0054	&	0.0031	&	0.0021	&	0.0046	&	0.0018	&	0.0024	&	0.0014	&	0.0030	&	0.0026	&	0.0015	&	0.0035	&	0.0015	&	0.0007	&	0.0040	\\
Algeria	&	0.0594	&	0.0352	&	0.0254	&	0.0592	&	0.0209	&	0.0279	&	0.0168	&	0.0348	&	0.0301	&	0.0171	&	0.0404	&	0.0177	&	0.0085	&	0.0473	\\
Andorra	&	0.0036	&	0.0010	&	0.0001	&	0.0001	&	0.0003	&	0.0003	&	0.0002	&	0.0004	&	0.0004	&	0.0002	&	0.0004	&	0.0002	&	0.0001	&	0.0006	\\
Angola	&	0.0240	&	0.0153	&	0.0128	&	0.0346	&	0.0098	&	0.0131	&	0.0079	&	0.0163	&	0.0141	&	0.0080	&	0.0189	&	0.0083	&	0.0040	&	0.0222	\\
Antigua	&	0.0008	&	0.0003	&	0.0001	&	0.0001	&	0.0001	&	0.0001	&	0.0001	&	0.0002	&	0.0002	&	0.0001	&	0.0002	&	0.0001	&	0.0000	&	0.0002	\\
Argentina	&	0.1940	&	0.0850	&	0.0335	&	0.0423	&	0.0374	&	0.0498	&	0.0302	&	0.0620	&	0.0539	&	0.0305	&	0.0723	&	0.0315	&	0.0151	&	0.0845	\\
Armenia	&	0.0024	&	0.0017	&	0.0017	&	0.0053	&	0.0012	&	0.0016	&	0.0009	&	0.0019	&	0.0017	&	0.0010	&	0.0023	&	0.0010	&	0.0005	&	0.0027	\\
\end{longtable}

\end{center}

\end{document}